\documentclass[hidelinks,journal,twocolumn,superscriptaddress,floatfix,footinbib,notitlepage,groupaddress,showpacs]{IEEEtran}

\usepackage{times,color,xcolor,amsfonts,amssymb,amsbsy,amsthm,amsmath,graphics,graphicx,hyperref,bm,bbm,makecell,multicol,multirow}
\usepackage{pdflscape}    
\usepackage{longtable}  

\newcommand{\ignor}[1]{}

\DeclareFontFamily{OT1}{pzc}{}
\DeclareFontShape{OT1}{pzc}{m}{it}%
{<-> s * [1.25] pzcmi7t}{}
\DeclareMathAlphabet{\mathpzc}{OT1}{pzc}%
{m}{it}

\let\oldsqrt\sqrt
\def\sqrt{\mathpalette\DHLhksqrt}
\def\DHLhksqrt#1#2{%
	\setbox0=\hbox{$#1\oldsqrt{#2\,}$}\dimen0=\ht0
	\advance\dimen0-0.2\ht0
	\setbox2=\hbox{\vrule height\ht0 depth -\dimen0}%
	{\box0\lower0.4pt\box2}}

\begin{document}
	\title{Quantum-Classical Access Networks with Embedded Optical Wireless Links}
	
	\author{Osama~Elmabrok,~\IEEEmembership{Student~Member,~IEEE,}
		 	  Masoud Ghalaii,~\IEEEmembership{Student~Member,~IEEE,}
				and Mohsen~Razavi 

		\thanks{This work was presented in part at the IEEE Globecom Conf. 2016 in Washington, DC. This research has partly been funded by the UK EPSRC Grants EP/M506953/1 and EP/M013472/1, the ministry of higher education and scientific research (MHESR) in Libya, and White Rose Research Studentship. The authors are with the School of Electronic and Electrical Engineering, University of Leeds, Leeds, LS2 9JT, UK (e-mail: elome@leeds.ac.uk, elmgh@leeds.ac.uk, and M.Razavi@leeds.ac.uk).}}
		
	\maketitle

\begin{abstract}

We examine the applicability of wireless indoor quantum key distribution (QKD) in hybrid quantum-classical networks. We propose practical configurations that would enable wireless access to such networks. The proposed setups would allow an indoor wireless user, equipped with a QKD-enabled mobile device, to communicate securely with a remote party on the other end of the access network. QKD signals, sent through wireless indoor channels, are combined with classical ones and sent over shared fiber links to the remote user. Dense wavelength-division multiplexing would enable the simultaneous transmission of quantum and classical signals over the same fiber. We consider the adverse effects of the background noise induced by Raman scattered light on the QKD receivers due to such an integration. In addition, we consider the loss and the background noise that arise from indoor environments. We consider a number of discrete and continuous-variable QKD protocols and study their performance in different scenarios.

\end{abstract}



\section{Introduction}

\IEEEPARstart{F}{uture} communications networks must offer improved security features against possible attacks enabled by quantum computing technologies. One possible solution is to develop quantum-classical networks that allow any two users to, not only exchange data, but also share secret key bits using quantum key distribution (QKD) techniques. Such a key can then be used to enable secure transmission of data between the two users. QKD technology is commercially available today~\cite{idquantique,QuantumCTek} and it has been used to exchange secret keys between pairs of users connected via fiber \cite{QKD_10Gbps_DWDM} or free space \cite{Zeilinger_Decoy_07} channels. QKD has also been implemented in several network settings~\cite{SECOQC2009, TokyoQKDNetwork2011, Chinanetworks2009}. Despite this progress, more work needs to be done to make QKD conveniently available to the end users of communications networks. In this paper, we address {\em wireless} access to a hybrid quantum-classical network. We consider hybrid links, with or without a trusted/untrusted relay point, between a wireless end user and the corresponding central office in an access network. This is done by adopting wireless indoor QKD links and embedding them into fiber-based passive optical networks (PONs).

QKD enables two remote users, Alice and Bob, to generate and exchange provably secure keys guaranteed by the laws of quantum physics~\cite{Gisin2002quantum, scarani2009security}. The obtained secret keys can then be used for data encryption and decryption between the two intended users. In conventional QKD protocols, an eavesdropper, Eve, cannot intercept the key without disturbing the system, and accordingly having her presence discovered. Furthermore, because of the no-cloning theorem~\cite{no-cloning}, Eve cannot exactly copy unknown quantum states. Based on these two principles, Bennett and Brassard in 1984 came up with their BB84 protocol in which {\em single photons} were carrying the key-bit information from Alice to Bob \cite{Bennett_BB84}. Over the time, more practical protocols have been developed that allow us to use weak laser pulses instead of ideal single-photon sources \cite{MXF:Practical:2005}. Nevertheless, most QKD protocols will still rely on the few-photon regime of operation, which makes them vulnerable to loss and background noise. This will make the implementation of QKD especially challenging in wireless mobile environments in which background noise is strong and alignment options are limited \cite{Wireless_indoor_QKD, Globecom15}. 

However challenging, embedding QKD capability into mobile/handheld devices is an attractive solution for exchanging sensitive data in a safe and convenient manner. For instance, customers in a bank can exchange secret keys wirelessly with access points in the branch without waiting for a teller or a cash machine. Initial prototypes have already been made, which enable a handheld device to exchange secret keys with an ATM without being affected by skimming frauds \cite{HP_HandheldQKD,chun2017handheld}. As another application, it would be desirable to enable a user working in a public space, such as an airport or a cafe, to exchange secret keys with its service provider via possibly untrusted nodes. Similarly, once fiber-to-the-home infrastructure is widely available, home users should benefit from such wireless links that connect them, via a PON, to other service provider nodes. In this case, the connection to the PON can be via an internal QKD node trusted by the user. Note that, in all cases above, we are dealing with a wireless link in an {\em indoor} environment, which may offer certain advantages, as compared to a general outdoor setup, in terms of ease of implementation. It is then a proper starting point for offering wireless QKD services as we study in this paper.

The above scenarios require hybrid links on which both data and quantum signals can travel in both wireless and wired modes. In this case, wireless QKD signals must somehow be collected and sent to the nearest service provider node over an optical fiber. In order to have a cost effective solution, the collected wireless QKD signals should be transmitted along with classical data signals over the same fiber links. A QKD system run on such a hybrid quantum-classical link would then face certain challenges. First, the background light in the wireless environment can sneak into the fiber system and increase error rates of the QKD setup. Furthermore, due to nonlinear effects in optical fibers such as four-wave mixing and Raman scattering~\cite{Eraerds2010_1Gbps}, the data channels that travel alongside QKD channels on the same fiber can induce additional background noise on QKD systems. In particular, the impact of the Raman scattered light can be severe \cite{Eraerds2010_1Gbps}, because its spectrum can overlap with the frequency band of QKD channels. By using extensive filtering in time and frequency domains, the impact of this noise can be mitigated~\cite{Patel2012coexistence, QKD_10Gbps_DWDM, Bahrani2016Crosstalk} and even maximally reduced~\cite{Bahrani2016orthogonal}, but it cannot be fully suppressed.

In this paper, by considering the effect of various sources of noise mentioned above, four setups for embedding wireless indoor QKD links into quantum-classical access networks are investigated. In each case, we find the corresponding key generation rate for relevant QKD protocols. We use the decoy-state BB84 (DS-BB84)~\cite{MXF:Practical:2005}, which relies on weak laser pulses, and measurement-device-independent QKD (MDI-QKD)~\cite{Lo2012MDI-QKD} protocols in our setups. The latter protocol can provide a trust-free link, as required in the case of a user in a public space, between the wireless user and the central office in an access network. The price to pay, however, is possible reduction in the rate. We also consider the GG02 protocol~\cite{GG02}, as a continuous-variable (CV) QKD scheme, and compare it with our discrete-variable (DV) protocols in terms of resilience to background noise  and loss~\cite{kumar2014experimental,lasota2017robustness}. CV QKD receivers require standard telecommunications technology for coherent detection, and in that sense they do not rely on single-photon detectors as their DV counterparts do.

The remainder of this paper is organized as follows. In Sec.~\ref{Sec:SystemDescription}, the system is described and in Sec.~\ref{Sec:KeyRateAnalysis} the key rate analysis is presented. The numerical results are discussed in Sec.~\ref{Sec:NumericalResults}, and Sec.~\ref{Sec:Conclusions} concludes the paper.

\section{System Description}
\label{Sec:SystemDescription}

\begin{figure}[t]
	\centering
	\includegraphics[width=.85\linewidth]{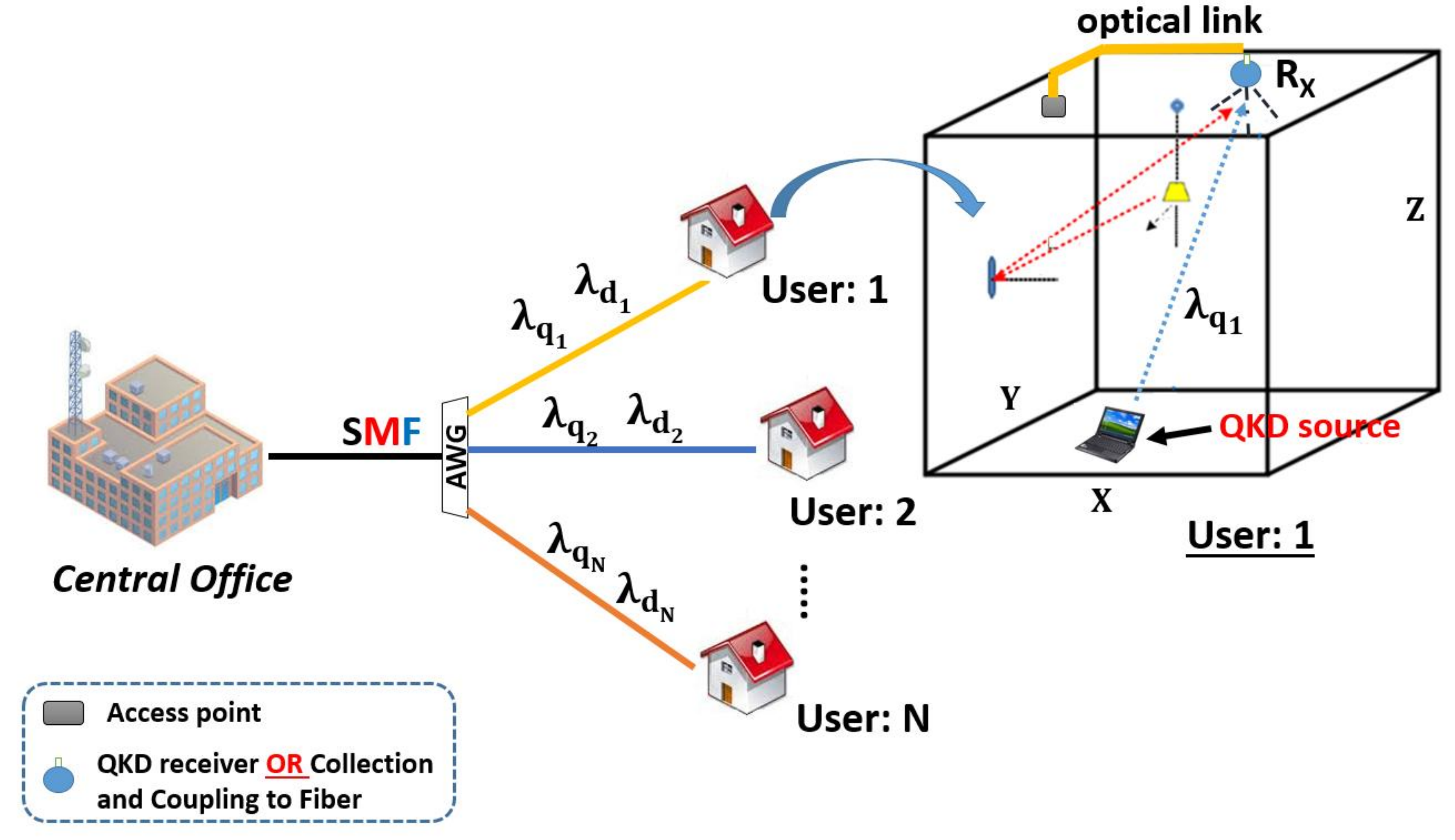}
	\caption{Schematic view of exchanging secret keys between an indoor wireless user with a central office at the end of an access network. The transmitter is mobile, while the QKD receiver or the collection point is fixed on the ceiling.}
	\label{fig_schematic_view}
\end{figure}

In this section, we describe our proposed setups for hybrid quantum-classical access networks comprised of optical wireless and fiber-optic links. Such setups can wirelessly connect a mobile user, in indoor environments, to the central office in access networks; see Fig.~\ref{fig_schematic_view}. We assume a total of $N$ end users, which are connected to the central office via a dense wavelength-division multiplexing (DWDM) PON. The corresponding wavelengths assigned to quantum and classical data channels are, respectively, denoted by $Q = \lbrace\lambda_{q_1}$, $\lambda_{q_2}$, ...,$\lambda_{q_N} \rbrace$ and $D$ = $\lbrace$$\lambda_{d_1}$, $\lambda_{d_2}$, ...,$\lambda_{d_N}$$\rbrace$. The $k$th user, $k = 1,\ldots, N$, employs wavelength $\lambda_{q_k}$ ($\lambda_{d_k}$) to communicate his/her quantum (classical) signals to the central office, as shown in Fig.~\ref{fig_schematic_view}. The same wavelengths are also used for the downlink. In order to heuristically reduce the Raman noise effect, we assume that the lower wavelength grid is allocated to the QKD channels, while the upper grid is assigned to data channels~\cite{Bahrani2016Crosstalk}. In principle, one can optimize the wavelength allocation such that the Raman noise on the quantum channels is minimized \cite{Bahrani_optimal}.  

For our wireless user, we consider a particular indoor environment, in which it has been shown that wireless QKD is feasible \cite{Wireless_indoor_QKD,Globecom15}. In this setting, a window-less room, of $X\times Y \times Z$ dimensions, is lit by an artificial light source. The possibly mobile QKD transmitter is placed on the floor and it transmits light toward the ceiling. The transmitter module may or may not be equipped with beam steering tools. In the former case, we assume that a minimal manual alignment is in place, by which the QKD source is facing the ceiling. This can be achieved by providing some instructions for the end user during the QKD protocol. The QKD receiver or the signal collector is fixed at the center of the room's ceiling; see Fig.~\ref{fig_schematic_view}. We assume that, by using some dynamic beam steering, maximum possible power is collected from the QKD source. This may be achieved by using additional beacon pulses. The collected light may go through a non-imaging optical concentrator, such as a compound parabolic collector, and then be filtered by a bandpass filter before being detected or sent out toward its final destination. 

In each setup, we particularly study three different cases regarding the position of the mobile QKD device. Case 1 refers to the scenario when the QKD transmitter is placed at the center of the room's floor and emits light upward with semi-angle at half power of $\Phi_{1/2}$. In case 2, the same QKD transmitter as in case 1 is moved to a corner of the room in order to assess the mobility features. These cases will represent the best and the worst case scenarios in terms of channel loss, when minimal beam alignment is used at the transmitter end. In case 3, the light beam at the QKD source is narrowed and is directed toward the QKD receiver or the coupling element. This would correspond to the worst case scenario when beam alignment is available at both the source and the receiver. In all cases, we assume a static channel in our analysis, that is we assume that the channel does not change during the key exchange procedure. The real mobile user is then expected to experience a quality of service bounded by the worst and best-case scenarios above. In the following, we first describe our proposed setups and the QKD protocols used in each case, followed by a description of the channel model.  

\subsection{The proposed setups}

We consider four setups in which an indoor wireless user, Alice, equipped with a QKD-enabled  mobile device, would exchange secret keys with a remote party, Bob, located at the central office. {In order to keep the mobile user's device simple, we assume that Alice is only equipped with the QKD encoder. That would imply that certain QKD schemes, such as entanglement-based QKD \cite{BBM_92}, are not suitable for our purpose if they require measuring single photons at the mobile user's end. Bob, however, represents the service provider node and could be equipped with the encoder and/or the decoder module as needed.} Based on these assumptions, here, we consider several settings depending on the existence or non-existence of a trusted/untrusted relay point between the wireless user and Bob at the central office. In all setups, a data channel will be wavelength multiplexed with the quantum one to be sent to the central office. We assume that classical data is being modulated at a constant rate throughout the QKD operation. 

\subsubsection{Setup 1 with a trusted relay point}

\begin{figure}[t]
	\centering
	\includegraphics[width=.90\linewidth]{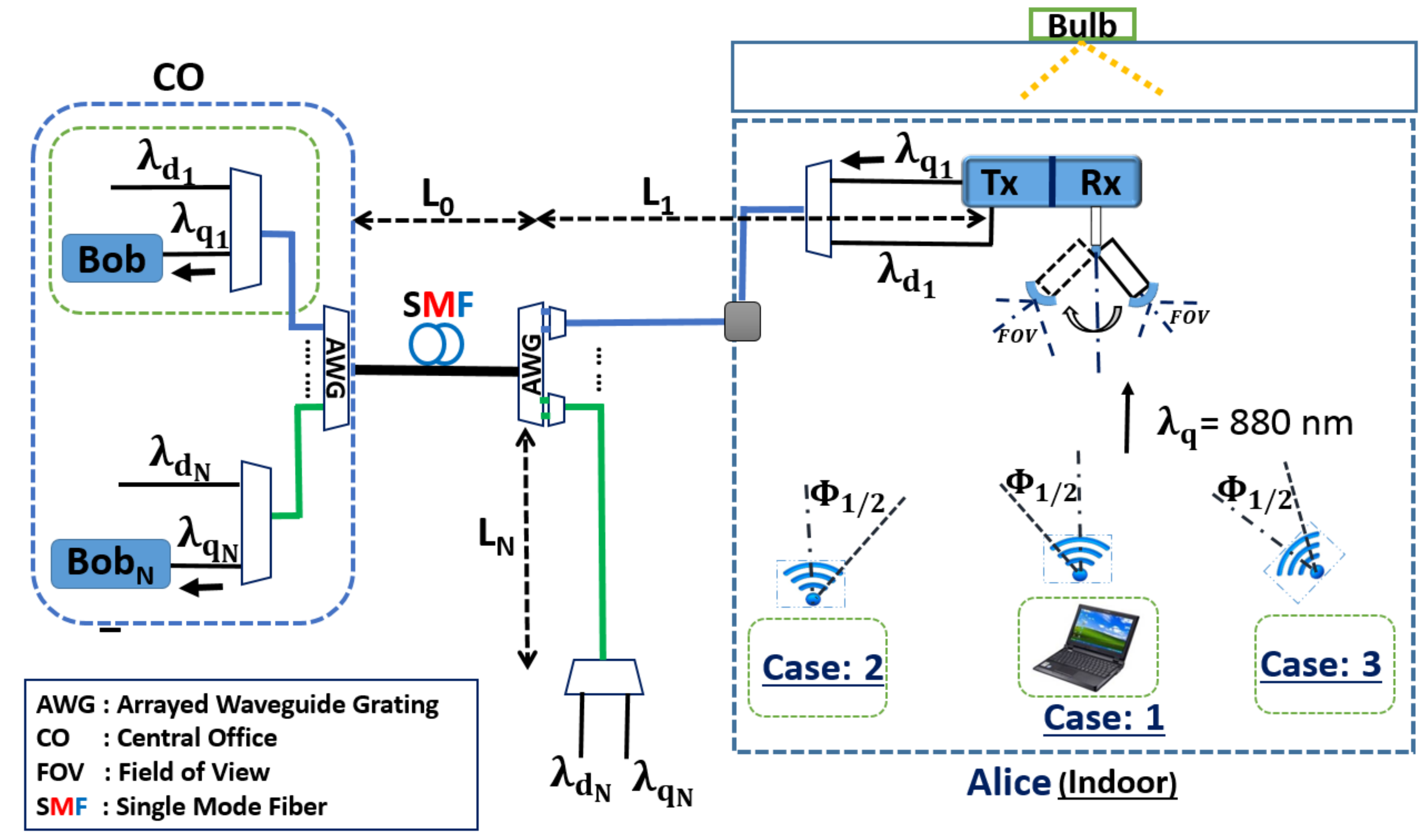}
	\caption{Setup 1, where secret key exchange between Alice and Bob is achieved in two steps. $K_1$ is generated between Alice and Rx, while $K_2$ is generated between Tx and Bob. The resultant key is computed by taking the XOR of $K_1$ and $K_2$. Three cases are examined according to the position and alignment of the QKD transmitter. The DS-BB84 and GG02 protocols will be examined in this setup. Dynamic beam steering is used at the Rx node.}
	\label{fig_scenario_1}
\end{figure}

Setup 1 is applicable whenever a trusted node between the sender and the recipient exists. For instance, in an office, we can physically secure a QKD relay node inside the building with which the wireless QKD users in the room can exchange secret keys. In Fig.~\ref{fig_scenario_1}, such a node is located on the ceiling and it is comprised of Rx and Tx boxes. In this setup, the secret key exchange between Alice and Bob is accomplished in two steps: a secret key, $K_1$, is generated between Alice and the Rx box in Fig.~\ref{fig_scenario_1}; also, independently but in parallel, another secure key, $K_2$, is exchanged between Tx and the relevant Bob in the central office. The final secret key is then obtained by applying an exclusive-OR (XOR) operation to $K_1$ and $K_2$. Note that in this setup both links are completely run separately; therefore, the wavelength used in the wireless link does not need to be the same as the wavelength used in the fiber link. In fact, for the wireless link, we use 880~nm range of wavelength, for which efficient and inexpensive single-photon detectors are available. For the fiber link, conventional telecom wavelengths are used. DS-BB84 and GG02 protocols will be used for this setup.


\subsubsection{Setup 2 without a relay point}

\begin{figure}[t]
	\centering
	\includegraphics[width=.90\linewidth]{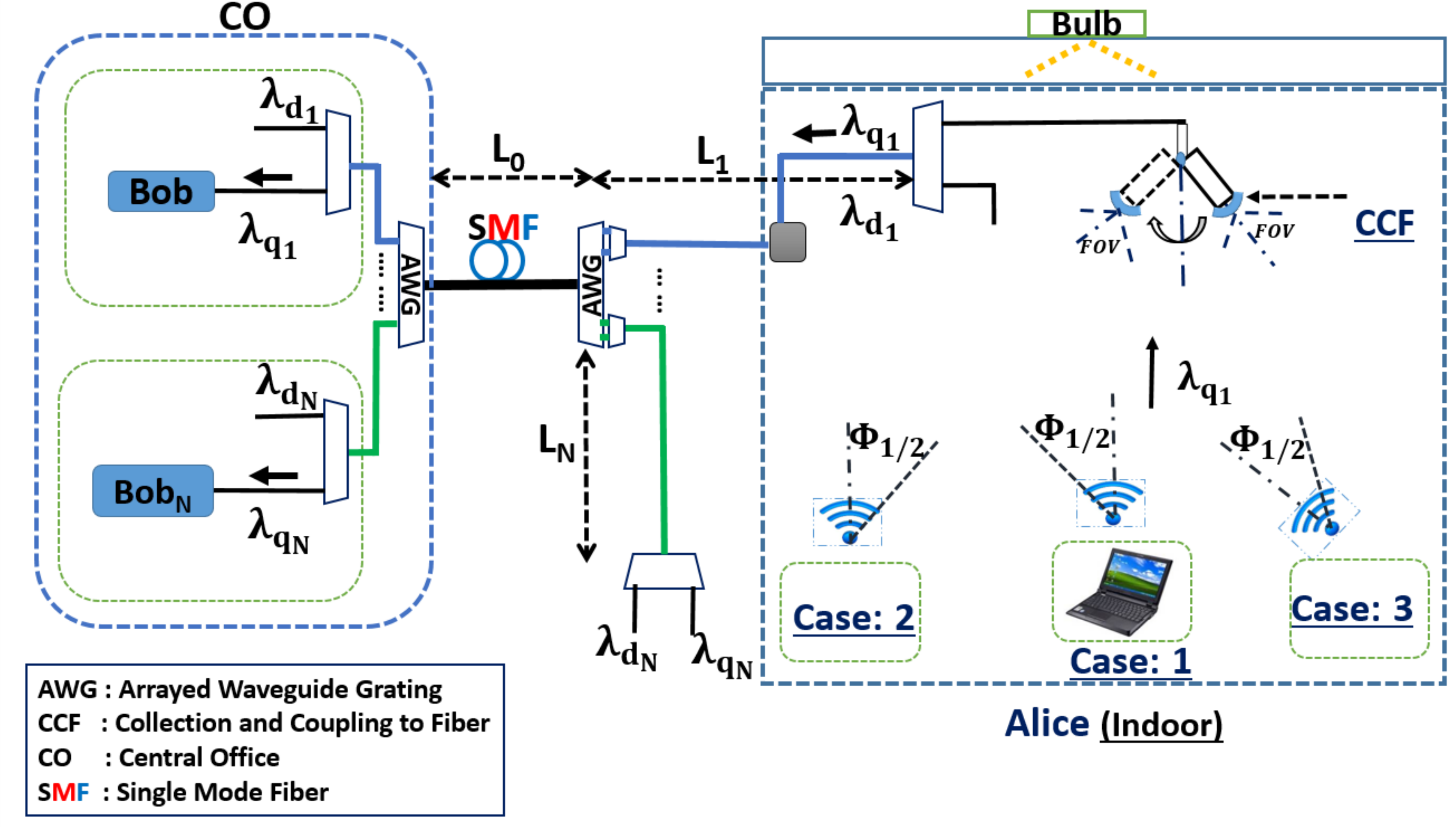}
	\caption{Setup 2, where secret keys are exchanged between Alice and Bob using the DS-BB84 and GG02 protocols. The latter is only used in case 3. The QKD signals are collected and coupled to the fiber and sent to Bob, where the measurement is performed. Dynamic beam steering is used at the collection node.}
	\label{fig_scenario_2}
\end{figure}

In this setup, we remove the need for having a relay point altogether. As shown in Fig.~\ref{fig_scenario_2}, the signals transmitted by Alice are collected by a telescope and coupled to a single-mode fiber to be sent to the central office. QKD measurements will then be performed at the central office. Because of this coupling requirement, the wireless signals undergo an additional coupling loss in setup 2. To reduce the coupling loss, in this setup, and, for fairness, in all others, we assume that the telescope at the collection point can focus on the QKD source. This can be achieved by additional beacon beams and micro-electro-mechanical based steering mirrors \cite{chun2017handheld}. In order to efficiently couple this photon  to the fiber, the effective FOV at the collection point should match the numerical aperture of a single-mode fiber. That requires us to use FOVs roughly below $6^\circ$, although, in practice, much lower values may be needed. In this setup, DS-BB84 and GG02 can be suitable protocols and will be examined in the following sections. 

\subsubsection{Setups 3 and 4 with untrusted relay points}	 

\begin{figure}[t]
	\centering
	\includegraphics[width=.90\linewidth]{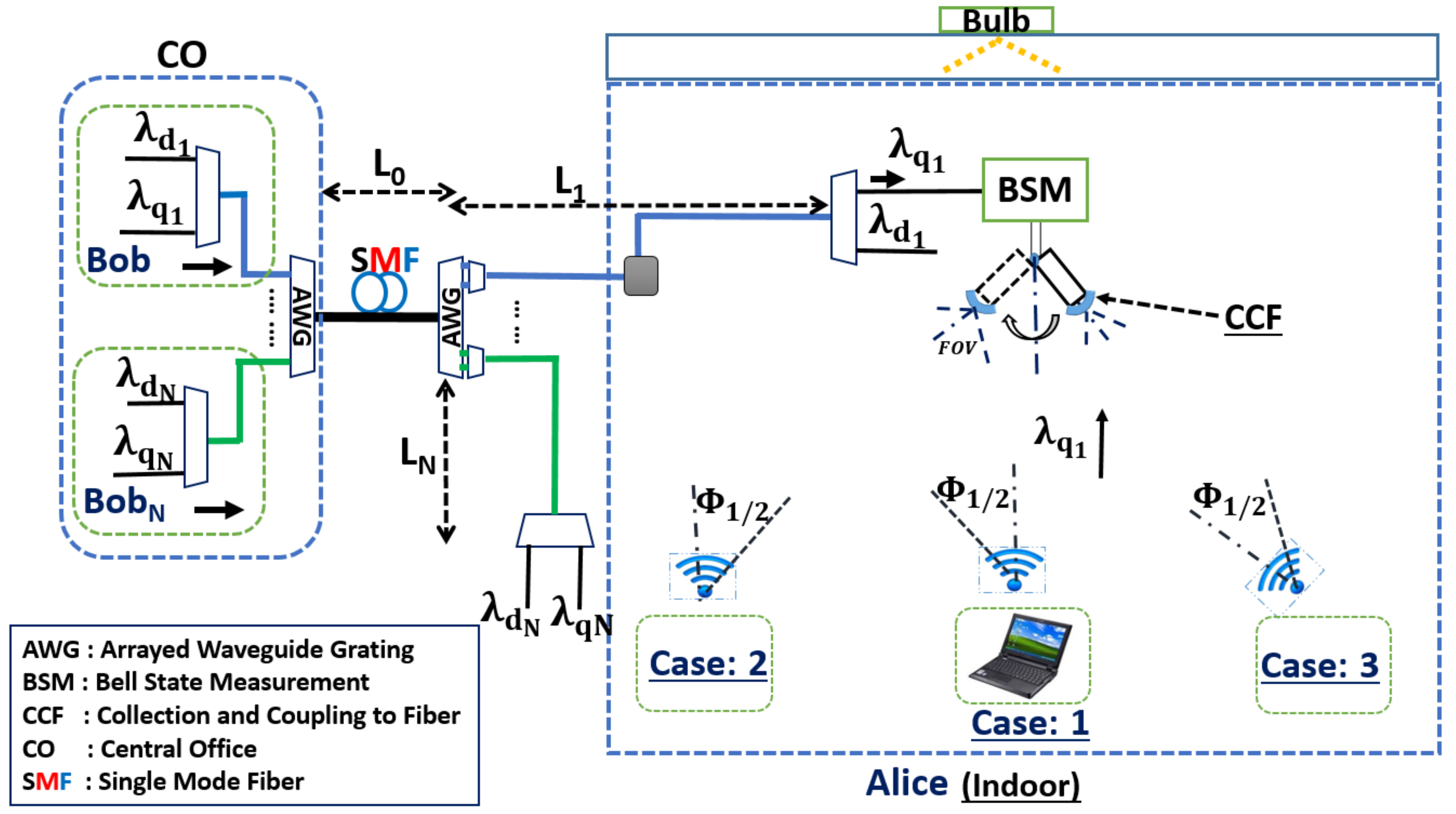}
	\caption{Setup 3, where secret keys are exchanged between Alice and Bob using the MDI-QKD protocol. The BSM is performed at the user's end in this setup.}
	\label{fig_scenario_3}
\end{figure}

\begin{figure}[t]
	\centering
	\includegraphics[width=.90\linewidth]{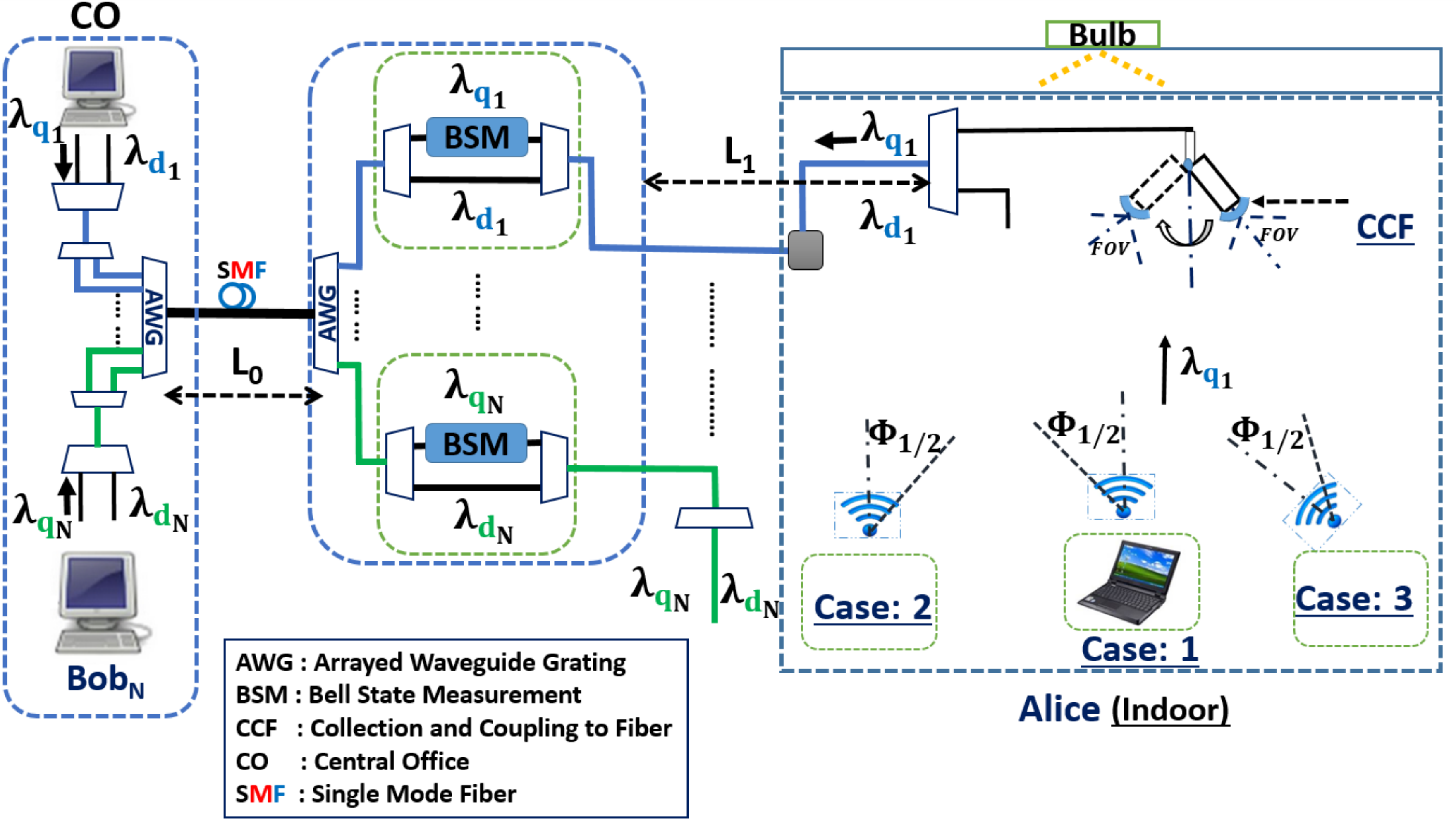}
	\caption{Setup 4, where secret keys are exchanged between Alice and the central office using the MDI-QKD protocol. The BSM is performed at the splitting point of the DWDM PON. }
	\label{fig_scenario_4}
\end{figure}

The setups in Figs.~\ref{fig_scenario_3} and \ref{fig_scenario_4} are of interest whenever the indoor environment that the wireless user is working at is not trustworthy. For instance, if the user is working at a public place, such as a coffee shop or an airport, s/he may not necessarily trust the owners of the local system. In such setups, we can use the MDI-QKD technique~\cite{Lo2012MDI-QKD} to directly interfere the quantum signal sent by the users with that of the central office. This can be accomplished by, if necessary, coupling the wireless signal into the fiber and performing a Bell-state measurement (BSM) on the photons sent by Alice and Bob at either the user's end (setup 3), or at a certain place located between the sender and the recipient at the central office (setup 4). In setup 4, we use the splitting terminal of a PON to implement such BSMs. Note that in setups 3 and 4 we need to interfere a single-mode signal traveling in fiber with a photon that has traveled through the indoor channel. In order to satisfy the BSM indistinguishability criterion, we then need to collect only one spatial mode from the wireless channel. The flexible beam steering used at the collection node should then satisfy this requirement. 

\begin{figure}[t]
	\centering
	\includegraphics[width=.85\linewidth]{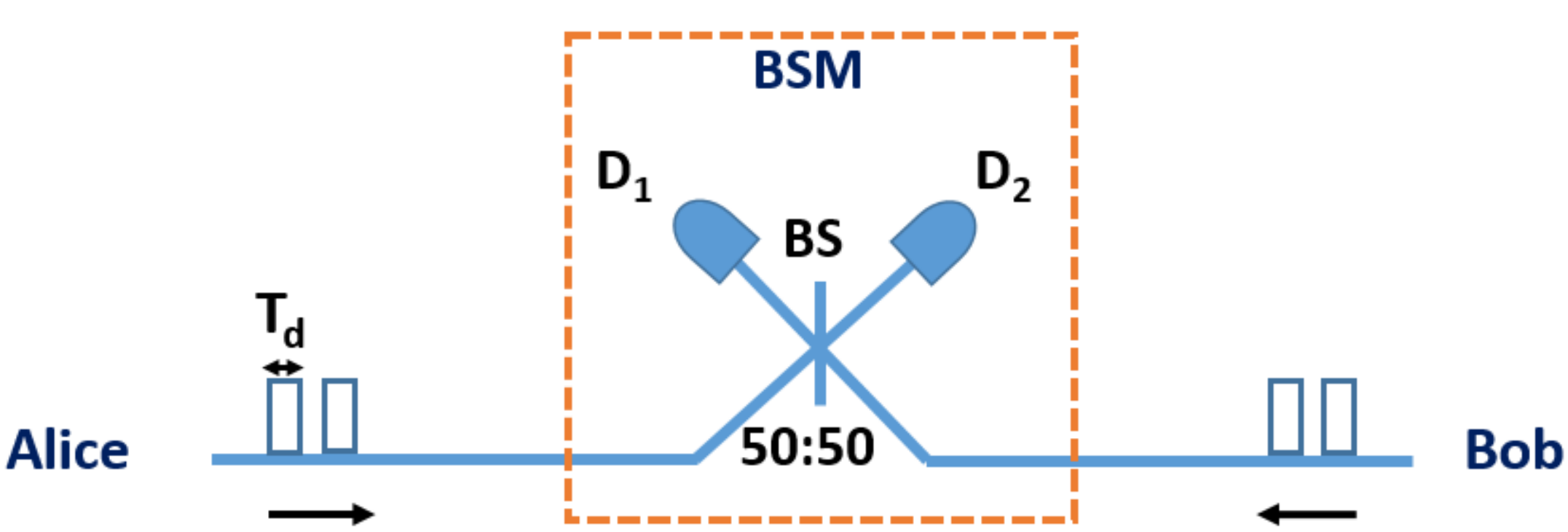}
	\caption{The Bell-state measurement (BSM) module used in setups 3 and 4. This module works for time-bin encoded QKD signals. If fast detectors are available, as assumed here, we can do a separate measurement on each time bin. If not, we can still measure one out of four Bell states by relying on a single click in total on each detector.}
	\label{fig_BSM_2}
\end{figure}

Here, we use a probabilistic setup for the BSM operation, as shown in Fig.~\ref{fig_BSM_2}. In this setup, the light coming from the two users are coupled at a 50:50 (fiber-based) beam splitter and then detect the outgoing signals using single-photon detectors. This simple setup is suitable for time-bin encoding techniques in QKD, which offer certain advantages in both fiber and free-space QKD systems. In particular, they may suffer less from alignment issues as compared to polarization-based encoding in wireless environments. Note that two successive clicks, one corresponding to each time bin, is required to have a successful BSM. That would require fast single-photon detectors with sub-nanosecond deadtimes. This is achievable using self-difference feedback techniques developed recently \cite{Yuan:Selfdif:2007}. If such detectors are not available, one can rely on one click on each detector, which roughly corresponds to declaring half of the success cases.

\subsection{The employed QKD protocols}

We use a number of discrete and continuous-variable QKD protocols to investigate the performance of the proposed configurations. In the case of DV protocols, we use the time-bin encoding, in which the information is encoded onto the phase difference between two successive pulses \cite{time-bin_encoding}. We assume that the gap between the two pulses is sufficiently short that similar phase distortions would be applied to both time bins while traversing the channel. Possible discrepancies are modeled by a relative-phase error term $e_d$. In the following, we provide a brief description of QKD protocols considered in this paper.

\subsubsection{DS-BB84}

In the ideal BB84 protocol~\cite{Bennett_BB84}, it is assumed that Alice, the sender, uses a single-photon source. However, this is not necessarily the case in practice. The actual alternative source is a highly attenuated laser that produces weak coherent states. The problem with using such sources is the possibility of experiencing the photon-number-splitting (PNS) attack~\cite{Brassard2000limitations} as each pulse might contain more than one photon. That is, Eve can siphon a photon and forward the rest to Bob. Later, after public announcement of the bases by Alice and Bob, Eve can measure exactly the state of the photon without revealing her presence. The decoy-state technique was proposed to beat this kind of attack \cite{hwang2003quantum}. The idea is to use several different light intensities, instead of one, so that any attempts by Eve to intrude on the link is more likely to be detected~\cite{MXF:Practical:2005}. In our key-rate analysis, we use the efficient version of DS-BB84~\cite{Lo2005efficient}, where $Z$ basis is chosen more frequently than the $X$ basis. In the time-bin encoding, the $Z$ basis is spanned by the single-photon states corresponding to each time-bin, whereas the $X$ eigenbases are the superposition of such states. We also assume that a passive Mach-Zehnder interferometer is used for decoding purposes.


\subsubsection{MDI-QKD}

The MDI-QKD protocol provides an efficient method of removing all detector side-channel attacks~\cite{Lo2012MDI-QKD}. This is done by performing the measurement by a third party, Charlie, who is not necessarily trusted. In this protocol, Charlie performs a BSM on Alice and Bob's signals, where each have a DS-BB84 time-bin encoder \cite{Ma2012alternative, MDIQKD_finite_PhysRevA2012}. Here, we again assume that the efficient version of DS-BB84 is in use. After Charlie announces the measurement outcomes of the successful events over a public channel, Alice and Bob follow the typical sifting and post processing procedures to come up with a shared secret key. 



\subsubsection{GG02}	  

While DV-QKD requires single-photon detectors, CV-QKD protocols are compatible with standard telecommunication technologies for coherent optical communications, namely, that of homodyne and heterodyne receivers~\cite{diamanti2015distributing}. CV-QKD has also, in certain regimes, the possible advantage of being more resilient to the background noise induced in WDM networks than DV-QKD \cite{CV_ResilienceBQi}. This is due to the intrinsic filtration of photons that do not match the spatio-temporal and polarization mode of the local oscillator (LO) in homodyne receivers \cite{kumar2014experimental}. However, for secure communication, CV-QKD may only be practical for short distances in comparison with DV-QKD~\cite{lodewyck2007quantum, jouguet2013experimental}. 
This is because of the excess noise and loss in the optical channels, as well as the limited efficiency of the classical reconciliation~\cite{Madsen2012entangledstates}.

The GG02 protocol is introduced by Grosshans and Grangier \cite{GG02}. It is the counterpart of the BB84 protocol in the CV prepare and measure schemes. In contrast to BB84, which relies on discrete variables, such as the polarization of single photons, GG02 exploits the quadratures of coherent states for encoding the information. In GG02, two random numbers, $X_A$ and $P_A$ are drawn by Alice according to two independent zero-mean Gaussian distributions with variance $V_A$, in the shot noise unit. The coherent state $|X_A+\textrm{i} P_A\rangle$ is then prepared, using amplitude and phase modulators, by Alice and sent to Bob, who randomly measures one of the two quadratures. After this stage both users acquire correlated random data. The error reconciliation and the privacy amplification are then performed in order to obtain the final secure key~\cite{GG02}. Reverse reconciliation \cite{grosshans2003quantum} is used in our study.

\subsection{Channel Characterization}

In this section, we model the two parts of our communication link, i.e., the wireless and fiber-based components and find out how much loss or background noise they may introduce.

\subsubsection{Indoor optical wireless channel}

A wireless QKD system ay suffer from two issues. The first is the existence of background noise caused by the artificial, as well as natural, sources of light in the room. The second important issue is the path loss, which can also have a severe impact on the QKD performance in indoor environments. The latter is modeled by the channel DC gain, $H_{\rm DC}$ \cite{kahn1997wireless,gfeller1979wireless}, which determines the portion of the transmitted power that will be detected at the receiver. 

In this paper, we follow the same methodology and assumptions, as presented in our recent work in ~\cite{Wireless_indoor_QKD,Globecom15}, to calculate the indoor channel transmittance, $H_{\rm DC}$ and the corresponding background noise. In our assumed window-less room, the background noise induced by the artificial lamp is calculated. That would depend on the power spectral density (PSD) of the employed light source. The receiver's FOV is also important since it limits the amount of background noise that may sneak into the QKD receiver. Here, we account for the reflected light from the walls and the floor that would be collected at the ceiling. We use optical wireless communication (OWC) models in~\cite{kahn1997wireless,gfeller1979wireless} for loss and background noise calculations. For the sake of brevity, we do not repeat that analysis here, but give some of the key relationships below. 

The DC-gain for a line-of-sight (LOS) link, which here is used to estimate the channel transmittance, is given by~\cite{kahn1997wireless}: 
\begin{align}
\label{eta_DC}
H_{\rm DC}= 
\begin{cases} 
\frac{A(m+1)}{2\pi d^2} \cos(\phi)^{m} T_s(\psi)  \\
\times  g(\psi) \cos(\psi)  ~~~~~~~~~~~~~~~~  0 \leq \psi \leq \Psi_c   \\ 
0 ~~~~~~~~~~~~~~~~~~~~~~~~~~~~~~~~~~ \text{elsewhere}, \end{cases}
\end{align}
where $d$ is the distance between the QKD sender and the QKD node on the ceiling; $\psi$ symbolizes the incidence angle with respect to the receiver axis, whereas $\phi$ represents the irradiance angle. Such parameters describe the relative position and orientation between the transmitter and receiver modules. For instance, the orientation in case 3 is modeled by assuming that the transmitter and receiver axes are identical and the beam angle is narrow. $T_s(\psi)$ is the filter signal transmission; $m$ and $g(\psi)$ are, respectively, the Lambert's mode number used to define the directivity of the source beam and the concentrator gain, which are given by
\begin{align}
m=\frac{-\ln 2}{\ln(\cos (\Phi_{1/2}))}
\end{align}
and
\begin{align}
\label{g_psi}
g(\psi)= \begin{cases} 
\frac{n^2}{\sin^2(\Psi_{c})} ~~~~~~ 0 \leq \psi \leq \Psi_c 
\\ 
0 ~~~~~~~~~~~~~~~~\psi > \Psi_c.
\end{cases}
\end{align}
In \eqref{eta_DC}-\eqref{g_psi}, $\Psi_c$, $\Phi_{1/2}$ and $n$ are, respectively, the receiver's FOV, semi-angle at half power of the light source and the refractive index of the concentrator. Note that the narrower the FOV, the higher the concentrator gain is. This, of course, meets certain practical constraints for very low FOVs, which we try to avoid.

\subsection{Optical fiber link}

As for the optical link, we make the following assumptions. We consider a loss coefficient $\alpha$ in dB/km in the single-mode fiber. We also assume that the loss contributed by each multi-port DWDM multiplexer, labeled as AWG (arrayed waveguide grating) in Figs.~\ref{fig_scenario_1}--\ref{fig_scenario_4} is $\Lambda$ in dB. We neglect the loss associated with two-to-one multiplexers.

As we mentioned earlier, the main source of background noise in QKD channels in a fiber link is Raman scattering. The Raman noise generated by a strong classical signal spans over a wide range of frequencies, hence can populate the QKD receivers with unwanted signals~\cite{Eraerds2010_1Gbps}. The receivers can be affected by forward and backward scattered light depending on their locations  and the direction of light propagation \cite{Bahrani2016orthogonal}. For a classical signal with intensity $I$ at wavelength $\lambda_d$, the power of Raman noise at a QKD receiver with bandwidth $\Delta \lambda$ centered at wavelength $\lambda_q$ is given by ~\cite{Eraerds2010_1Gbps,Patel2012coexistence}  
\begin{align}
{I^{f}_{R}}(I,L,\lambda_d,\lambda_q)=Ie^{-\alpha L}L \Gamma (\lambda_d,\lambda_q) \Delta \lambda 
\end{align}
for forward scattering, and 
\begin{align}
{I^{b}_{R}}(I,L,\lambda_d,\lambda_q)=I\frac{(1-e^{-2 \alpha L })}{2 \alpha} \Gamma (\lambda_d,\lambda_q) \Delta \lambda
\end{align}
for backward scattering, where $L$ is the fiber length and $\Gamma (\lambda_d,\lambda_q)$ is the Raman cross section (per unit of fiber length and bandwidth), which can be measured experimentally. In our work, we have used the results reported in~\cite{Eraerds2010_1Gbps} for $\lambda_d = 1550$~nm and have used the prescription in~\cite{Bahrani2016orthogonal} to adapt it to any other wavelengths in the C band. The transmitted power $I$ is also set to secure a bit error rate (BER) of no more than 10$^{-9}$ for all data channels.
A photodetector would then collect a total average number of photons, due to forward and backward scattering, respectively, given by 
\begin{align}
\label{murf}
{\mu^{f}_{R}}=\frac{\eta_d {I^{f}_{R}} \lambda_q T_d}{hc}
\end{align}
and
\begin{align}
\label{murb}
{\mu^{b}_{R}}=\frac{\eta_d {I^{b}_{R}} \lambda_q T_d}{hc},
\end{align}
where $T_d$, $\eta_d$ and $h$, respectively, represent the detectors' gate duration, their quantum efficiency and Planck's constant with $c$ being the speed of light in the vacuum.

\section{Key Rate Analysis}
\label{Sec:KeyRateAnalysis}
In this section, the secret key rate analysis for our proposed setups is presented considering non-idealities in the system. The secret key rate is defined as the asymptotic ratio between the number of secure bits and sifted bits. Without loss of generality, we only calculate the rate for user 1 assuming that there is no eavesdropper present. The DS-BB84~\cite{MXF:Practical:2005} and GG02 protocols are used for setups 1 and 2, while the MDI-QKD protocol~\cite{Lo2012MDI-QKD, Ma2012alternative} is employed for setups 3 and 4. 

\subsection{Setups 1 and 2}

\subsubsection{DS-BB84 protocol}

The lower bound for the key generation rate in the limit of an infinitely long key is given by~\cite{MXF:Practical:2005}
\begin{align}
\label{KeyRate_limit}
R\geq q\lbrace -Q_\mu f h(E_\mu)+Q_1[1-h(e_1)]\rbrace, 
\end{align}
where all new parameters are defined in Appendix~\ref{App:DS-BB84_KeyRate}. There, we show that the expected value for these parameters in our loss and background induced model for the channel mainly depends on two parameters: the overall efficiency of each link $\eta$, and the total background noise per detector, denoted by $n_N$. Here, $n_N$ accounts for both dark counts and background noise in the link. In the following, we specify how these parameters can be calculated in each setup. 

In setup 1, we have two links, a wireless link and a wired link. Below, the parameter values for each link will be calculated separately.  

\noindent{\bf Setup 1, wireless link:} For the wireless channel, we assume that the background noise due to the artificial lighting source is denoted by $n_{B_1}$, which can be calculated using the methodology proposed in \cite{Wireless_indoor_QKD}. In our calculations, we upper bound $n_{B_1}$ by considering the case where the QKD receiver is focused on the center of the room. The total noise per detector, $n_N$, is then given by
\begin{align}
{n_N=n_{B_1} \eta_{d_1}/2 + n_{dc}},
\end{align}
where $\eta_{d_1}$ is the detector efficiency, for the detector in the Rx box, and $n_{dc}$ is the dark count rate per pulse for each detector in the Rx box in Fig.~\ref{fig_scenario_1}. We neglect the impact of the ambient noise in our windowless room~\cite{Wireless_indoor_QKD}. The total transmissivity is also given by $\eta = H_{\rm DC} \eta_{d_1}/2$. The factor 1/2 represents the loss in the passive time-bin decoder consisted of a Mach-Zehnder interferometer.

\noindent{\bf Setup 1, fiber link:} As for the fiber-based link, the background noise is mainly induced by the Raman scattered light. In this setup, where Bob's receiver is at the central office, forward scattered light is generated because of the classical signals sent by the users and backward scattered light is due to the signals sent by the central office. The total power of Raman noise, at wavelength $\lambda_{q_1}$, for forward and backward scattering are, respectively, given by 
\begin{align}
I^{f}_{T1} =& [I_R^f(I,L_0 + L_1,\lambda_{d_1},\lambda_{q_1}) \nonumber \\
&+ \sum_{k=2}^N{I_R^f(Ie^{-\alpha L_k},L_0,\lambda_{d_k},\lambda_{q_1})}] 10^{-2\Lambda/10}
\end{align}
and
\begin{align}
I^{b}_{T1} = & [I_R^b(I,L_0 + L_1,\lambda_{d_1},\lambda_{q_1}) 
\nonumber \\
&+ \sum_{k=2}^N{I_R^b(I,L_0,\lambda_{d_k},\lambda_{q_1})}] 10^{-2\Lambda/10},
\end{align}
where $L_0$ is the total distance between the central office and the AWG box at the users' splitting point and $L_k$ is the distance of the $k$th user to the same AWG in the access network. In the above equations, we have neglected the out-of-band Raman noise that will be filtered by relevant multiplexers in our setup. For instance, in calculating $I_{T1}^f$, we account for the effect of the forward Raman noise by the data signal generated by User 1 over a total distance of $L_0+L_1$, but, a similar effect by other users is only accounted for over a distance $L_0$. That is because the AWG box filters most of the Raman noise at $\lambda_{q_1}$ generated over distances $L_k$ and their effect can be neglected. By substituting the above equations in \eqref{murf} and \eqref{murb}, the total background noise per detector, at Bob's end in Fig.~\ref{fig_scenario_1}, is given by
\begin{align}
n_{N}=\frac{\eta_{d_2} \lambda_{q_1}T_d}{2hc}(I^{f}_{T1}+I^{b}_{T1})+n_{dc},
\end{align}
where $\eta_{d_2}$ is the detector efficiency at Bob's receiver. Note that in setup 1 we consider two different values for $\eta_{d_1}$ and $\eta_{d_2}$.The reason is that the former corresponds to the avalanche photodiode (APD) single-photon detectors at 880~nm, while the latter could be for InGaAs APD single-photon detectors within the 1550~nm band. 

The total transmissivity $\eta$ for the fiber link is given by $\eta_{\rm fib}\eta_{d_2}/2$, where $\eta_{\rm fib}$ is the optical fiber channel transmittance including the loss associated with AWGs given by 
\begin{align}
\label{fiber_channel_transmittanc}
{\eta_{\rm fib}=10^{-[\alpha(L_1+L_0)+2\Lambda]/10}.}
\end{align}


\noindent{\bf Setup 2:} In setup 2, the total Raman noise power for forward and backward scattering, denoted by $I^{f}_{T2}$ and $I^{b}_{T2}$ are given by $I^{f}_{T1}$ and $I^{b}_{T1}$, respectively. The total background noise per detector at Bob's end in Fig.~\ref{fig_scenario_2} is then given by 
\begin{align}
\label{tot_backgroundNoise}
n_N = \frac{\eta_{d_2}}{2}\left[\frac{ \lambda_{q_1} T_d}{hc} \left( {I^{f}_{T2}} + {I^{b}_{T2}}\right)  + n_{B_1}\eta_{\rm fib} \eta_{\rm coup}\right] 
+ n_{dc},
\end{align}
where $\eta_{\rm coup}$ is the additional air-to-fiber coupling loss that the indoor background photons, generated by the bulb, will experience before reaching the QKD receiver. The total channel transmittance between the sender and the recipient in this setup is given by $\eta=H_{\rm DC}\eta_{\rm coup}\eta_{\rm fib}\eta_{d_2}/2$.

\subsubsection{GG02 protocol}

The secure key rate for GG02 with reverse reconciliation under collective attacks is given by~\cite{fossier2009field}
\begin{align}
K= \beta I_{AB}-\chi_{BE}, 
\end{align}
where $\beta$ is the reconciliation efficiency. $I_{AB}$ and $\chi_{BE}$ are, respectively, the mutual information between Alice and Bob, and the amount of information obtained by the adversary in reverse reconciliation. More details can be found in Appendix \ref{App:GG02_KeyRate}. 

GG02 is characterized by the channel transmissivity $\eta_{\rm ch}$ and the excess noise $\varepsilon$. For estimating the latter, we need to consider the contribution of the bulb, $\varepsilon_{b}$, as well as the Raman scattering, $\varepsilon_{r}$. The total excess noise, $\varepsilon$, is then given by $\varepsilon_{b} + \varepsilon_{r} + \varepsilon_{q}$, where $\varepsilon_{q}$ is any other additional noise observed in the experiment. In the Appendix~\ref{App:GG02_KeyRate} formulation, the excess noise terms must be calculated at the input. For chaotic sources of light, if the average noise count at the end of a channel with transmissivity $\eta_t$ is given by $n$, the corresponding excess noise at the input would be given by $2n/\eta_t$ \cite{Feasibility2010dense,kumar2015coexistence}. Below, we use this expression to calculate $\varepsilon_{b}$ and $\varepsilon_{r}$ assuming that both the Raman noise and the bulb-induced background noise are of chaotic-light nature.

\noindent{\bf Setup 1, wireless link:} In setup 1, the background noise due to the bulb is denoted by $n_{B_1}$. This is the total background noise at the Rx box input. Given that the LO would pick a single spatio-temporal mode with matching polarization, the corresponding count that sneaks into the homodyne receiver would be $n_{B_1}/2$. The corresponding excess noise would then be given by $\varepsilon_{b} = n_{B_1}/H_{\rm DC}$ and $\varepsilon = \varepsilon_{b} + \varepsilon_{q} $. In this case, $\eta_{\rm ch} = H_{\rm DC}$. In an experiment, $\varepsilon_{q}$ is often calculated by measuring the corresponding parameter, $\varepsilon_{q}^{\rm rec}$, at the receiver. In this case, $\varepsilon_{q} = \varepsilon_{q}^{\rm rec}/(\eta_{\rm ch} \eta_B)$, where $\eta_B$ is Bob's receiver overall efficiency.

\noindent{\bf Setup 1, fiber link:} In this case, $\eta_{\rm ch} = \eta_{\rm fib}$, $\varepsilon_{b} = 0$, and $\varepsilon_{r} = n_r/\eta_{\rm ch}$, where
\begin{align}
n_{r}=\frac{\lambda_{q_1}T_d}{hc}(I^{f}_{T1}+I^{b}_{T1}).
\end{align}

\noindent {\bf Setup 2:} In setup 2, $\eta_{\rm ch} = H_{\rm DC} \eta_{\rm coup} \eta_{\rm fib}$, $\varepsilon_{b} = n_{B_1}/H_{\rm DC}$, and $\varepsilon_{r} = n_r/\eta_{\rm ch}$, where
\begin{equation}
n_r = \frac{\lambda_{q_1} T_d}{hc} \left( {I^{f}_{T2}} + {I^{b}_{T2}}\right).
\end{equation}

In all CV-QKD setups, we assume that a phase reference for the LO is available at the receiver.

\subsection{Setups 3 and 4 with MDI-QKD protocol}

The secret key rate for the MDI-QKD setup is given in Appendix \ref{App:MDI-QKD}. The key parameters to find for this scheme are $\eta_a$ and $\eta_b$, which, respectively, correspond to the total transmissivity seen by Alice and Bob channels, as well as $n_N$, which is the total background noise per detector. Here we find these parameters for Setups 3 and 4.


\noindent{\bf Setup 3:}   
The total forward and backward Raman noise power for setup 3 at wavelength $\lambda_{q_1}$ are, respectively, given by
\begin{align}
I^{f}_{T3} =& [I_R^f(I,L_0 + L_1,\lambda_{d_1},\lambda_{q_1}) \nonumber \\
&+ e^{-\alpha L_1}\sum_{k=2}^N{I_R^f(I,L_0,\lambda_{d_k},\lambda_{q_1})}] 10^{-2\Lambda/10}, \nonumber \\
I^{b}_{T3} =& [I_R^b(I,L_0 + L_1,\lambda_{d_1},\lambda_{q_1}) 
\nonumber \\
&+ e^{-\alpha L_1} \sum_{k=2}^N{I_R^b(I e^{-\alpha L_k},L_0,\lambda_{d_k},\lambda_{q_1})}] 10^{-2\Lambda/10}. 
\end{align}
The total noise per detector, $n_N$, for setup 3 is then given by 
\begin{align}
n_N = \frac{\eta_{d_2}}{4}\left[\frac{ \lambda_{q_1} T_d}{hc} \left( {I^{f}_{T3}} + {I^{b}_{T3}}\right)  + n_{B_1} \eta_{\rm coup}\right] 
+ n_{dc},
\end{align}
where we account for one particular polarization entering the BSM module. 

In setup 3, $\eta_{a} = H_{\rm DC} \eta_{d2}\eta_{\rm coup}/2$ and $\eta_{b} = \eta_{d2}\eta_{\rm fib}/2$, assuming an average loss factor of 1/2 for polarization mismatch. Note that the two modes interfering at the BSM must have matching polarizations. This can be achieved passively by using polarization filters before the 50:50 beam splitter in the BSM, in which case, an average loss of 1/2 is expected, or, alternatively, we need to use active polarization stabilizer, for which the corresponding loss factor approaches one.


\noindent{\bf Setup 4:}
The total forward and backward Raman noise power for setup 4 at wavelength $\lambda_{q_1}$ are, respectively, given by
\begin{align}
I^{f}_{T4}=& [I_R^f(I,L_0,\lambda_{d_1},\lambda_{q_1})+ \sum_{k=2}^N{I_R^f(I,L_0,\lambda_{d_k},\lambda_{q_1})}] \nonumber \\
&\times 10^{-2\Lambda/10}+I_R^f(I,L_1,\lambda_{d_1},\lambda_{q_1}), \nonumber \\
I^{b}_{T4} = & [I_R^b(Ie^{-\alpha L_1},L_0,\lambda_{d_1},\lambda_{q_1}) \nonumber \\  & + \sum_{k=2}^N{I_R^b(Ie^{-\alpha L_k},L_0,\lambda_{d_k},\lambda_{q_1})}\nonumber \\
& +I_R^b(Ie^{-\alpha L_0},L_1,\lambda_{d_1},\lambda_{q_1})]
10^{-2\Lambda/10}.
\end{align}

The total noise per detector, $n_N$, for setup 4 is as follows
\begin{align}
n_N = \frac{\eta_{d_2}}{4}\left[\frac{\lambda_{q_1} T_d}{hc} \left( {I^{f}_{T4}} + {I^{b}_{T4}}\right)  + n_{B_1} \eta_{\rm coup}10^{-\alpha L_1/10}\right] 
+ n_{dc}.
\end{align}

In setup 4, $\eta_{a} =  H_{\rm DC} \eta_{d_2} \eta_{\rm coup} 10^{-\alpha L_1/10}/2$ and $\eta_{b} = \eta_{d_2} 10^{-[\alpha L_0+2\Lambda]/10}/2$.

\section{Numerical Results}
\label{Sec:NumericalResults}

\begin{table}[t]
	\caption{Nominal values used for our system parameters.}      
	\begin{tabular}{|c|c|} \hline
		{\bf System Parameters} & {\bf Nominal value}\\ \hline
		Number of users, $N$ & 32 \\
		Fiber attenuation coefficient, $\alpha$  & 0.2 dB/km \\
		AWG insertion loss, $\Lambda$ & 2 dB \\
		Room size, $X$,$Y$,$Z$ & ($4 \times 4 \times 3$) m$^3$ \\
		Semi-angle at half power of the bulb & $70^{\circ}$ \\
		Reflection coefficients of the walls and floor &  $0.7$  \\
		Detector area & $1$ cm$^2$     \\
		Refractive index of the concentrator&  1.5 \\
		Semi-angle at half power of QKD source, $\Phi_{1/2}$ & $20^{\circ}$, $1^{\circ}$ \\
		\hline
		{\bf DV-QKD Parameters} & {\bf Nominal value}\\
		\hline
		Average number of photons per signal pulse, $\mu=\nu$  & 0.5\\
		Error correction inefficiency, $f$   & 1.16 \\
		Dark count per pulse, $n_{dc}$   & $10^{-7}$ \\ 
		Detector gate width, $T_{d}$   & 100 ps \\  
		Relative-phase error probability, $e_{d}$   & 0.033 \\    
		Quantum efficiency of detector, $\eta_{d1}$,
		{at 880~nm} & 0.6 \\
		Quantum efficiency of detector, $\eta_{d2}$,
		{at 1550~nm}& 0.3 \\ 
		\hline
		{\bf CV-QKD Parameters} & {\bf Nominal value}\\
		\hline
		Reconciliation efficiency, $\beta$   & 0.95 \\  
		Receiver overall efficiency, $\eta_{B}$   & 0.6 \\ 
		Electronic noise (shot noise units), $v_{elec}$  & 0.015 \\
		Excess noise (shot noise units), $\varepsilon_q^{\rm rec}$  & 0.002 \\\hline
	\end{tabular}
	\label{Table} 
\end{table}

In this section, we provide some numerical results for secret key rates in the four proposed setups. We use a DWDM scheme with 100~GHz channel spacing in the C-band with 32 users. We define $Q$ = $\lbrace$1530.8 nm, 1531.6 nm,...,1555.62 nm$\rbrace$ and $D$ = $\lbrace$1560.4 nm, 1561.2 nm,...,1585.2 nm$\rbrace$ for quantum and classical channels, respectively. We assume that $\lambda_{q_1}$ is 1555.62 nm and the corresponding $\lambda_{d_1}$ is 1585.2 nm. The classical data is transmitted with launch power $I = 10^{(-3.85+\alpha L/10+2\Lambda/10)}$ mW, which corresponds to receiver sensitivity of -38.5 dB guaranteeing a BER of $\textless$ 10$^{-9}$~\cite{Bahrani2016orthogonal}. In all setups, we assume that $L_1= L_2= \cdots=L_N$ all equal to 500~m.

Other nominal parameter values used in our simulation are summarized in Table~\ref{Table}. These are based on values that are technologically available today. {In particular, for DV-QKD systems, we assume silicon-based single-photon detectors are used in the 800~nm regime (setup 1, indoor channel), whereas GaAs detectors may need to be used in the 1550~nm regime (all other setups). The former often have higher quantum efficiencies than the latter. That is why in our numerical parameters, $\eta_{d1}$ is twice as big as $\eta_{d2}$. The dark count rate in such detectors varies from (100--1000)/s for an APD, to (1--100)/s for superconducting detectors~\cite{DUSEK2006381}. The average dark count rate considered here is 1000/s, which, over a period of $100$~ps, will result in $n_{dc} = 10^{-7}$. In the CV-QKD system, $\eta_{B}$ is Bob's receiver overall efficiency, which includes detector efficiencies and any insertion loss in the homodyne receiver. The parameter $\beta$ is the efficiency of our post-processing, which nowadays exceeds 95\% \cite{PhysRevA.84.062317}. The parameter values chosen for the receiver electronic noise and excess noise correspond to the observed values in recent CV-QKD experiments \cite{jouguet2013experimental}. Based on the values chosen for our system parameters, relevant parameters in Sec.~\ref{Sec:KeyRateAnalysis}, such as $\eta_{\rm fib}$ and $\eta_{\rm ch}$, can be calculated from which parameter $\eta$ for each setup is obtained. The noise parameter $n_N$, for each setup, can similarly be found. The Raman noise terms, in particular, have been calculated by extracting the Raman cross section from the experimental measurements reported in \cite{Eraerds2010_1Gbps}. Note that, in our numerical calculations, we often vary the coupling loss to study system performance.}

In each setup, three cases are considered for the light beam orientation of the QKD source. In the first case, the semi-angle at half power of the QKD source is $\Phi_{1/2}$ = 20$^{\circ}$ while the QKD source is placed at the center of the room's floor. With the same $\Phi_{1/2}$, the QKD source is moved to the corner of the room in the second case. We use $\Phi_{1/2}$ = 1$^{\circ}$ in the third case where the QKD source is located at the corner of the room, as in the second case, but the beam is directed and focused toward the QKD receiver or the collection element. A full alignment is assumed in the third case, while in the other two cases the QKD source is sending light upward to the ceiling with a wider beam angle.  As for the receiver,  we assume that its telescope is dynamically rotating to collect the maximum power from the user in the three cases. We assume that the effective receiver's FOV would correspond to the numerical aperture (NA) of a single-mode fiber. For single-mode fibers, NA is about 0.1, which means that the corresponding FOV that can be coupled to the fiber is around 6$^{\circ}$. Here, the QKD receiver's FOV is assumed to be 6$^{\circ}$ in order to maximize the collected power.

\begin{figure}[h]
	\centering
	\includegraphics[width=.95\linewidth]{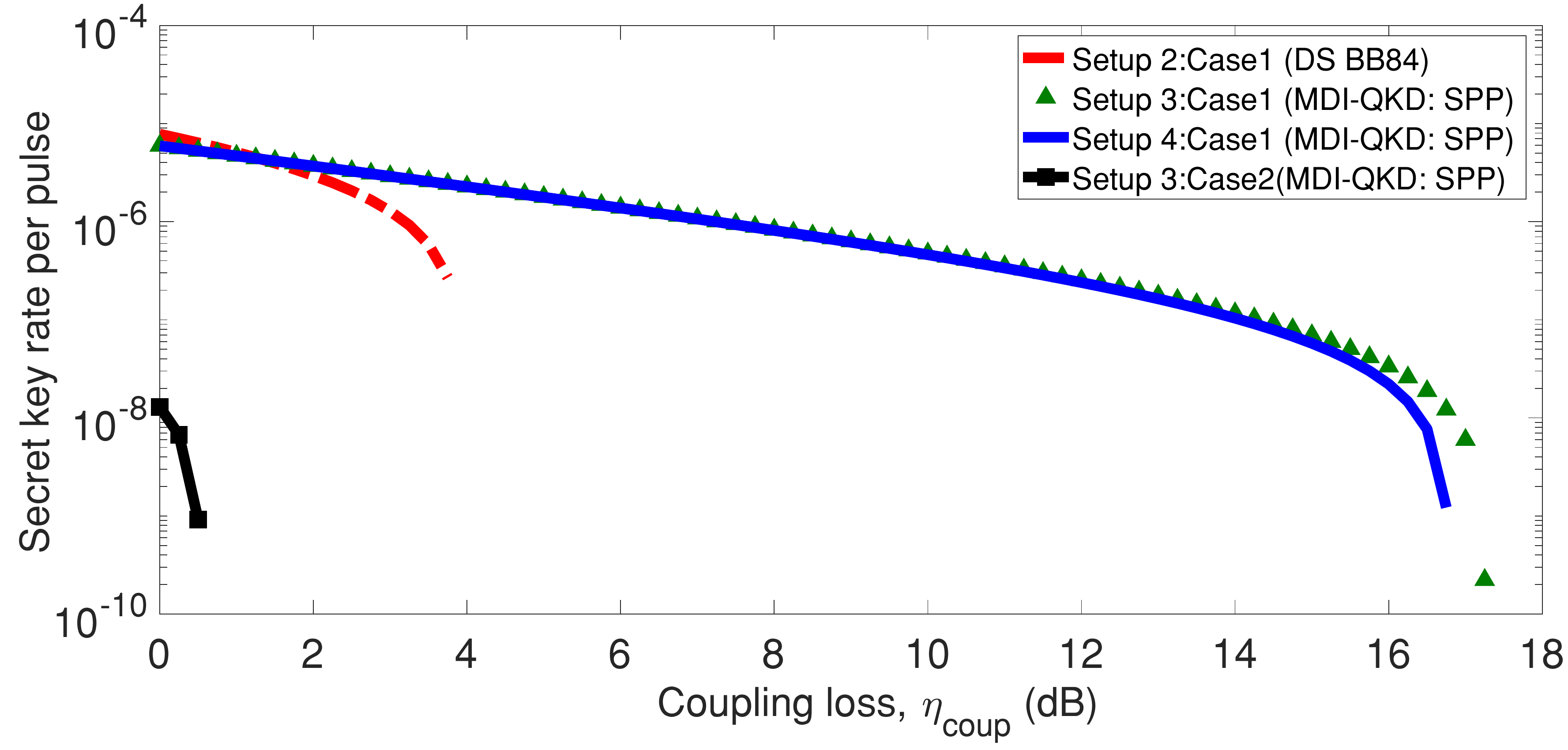}
	\caption{The secret key rate per pulse versus the coupling loss, $\eta_{\rm coup}$, in dB, in setups 2, 3 and 4 in cases 1 and 2. The QKD source is placed at the center of the room in case 1, while it is moved to a corner of the room in case 2, with semi-angle at half power of $\Phi_{1/2}$ = 20$^{\circ}$ in both cases. Receiver's FOV is 6$^{\circ}$. The decoy-state and MDI-QKD protocols are used for secret key rate analysis. The bulb's PSD in cases 1 and 2 is $10^{-7}$ W/nm and $10^{-8}$ W/nm, respectively. The fiber length ($L_0$) is 10 km. (DS: Decoy state; SPP: Single-photon pulse.)}
	\label{fig_rate_vs_couping_loss_case1_2}
\end{figure}

The first thing we study here is whether the loose alignment in cases 1 and 2 would be sufficient for the proper operation of a networked wireless link. The short answer turns out to be negative for setups 2--4. We already know the result for setup 1 from the previous work in \cite{Wireless_indoor_QKD}, in which the authors show that, if the only source of lighting in the room is an LED bulb with a PSD on the order of $10^{-5}$--$10^{-6}$~W/nm, then there will be regions over which even in cases 1 and 2 the wireless user can exchange secret keys with the Rx box. This seems to no longer necessarily hold if we remove the trusted relay node in the room. In Fig.~\ref{fig_rate_vs_couping_loss_case1_2}, we have plotted the secret key rate versus the coupling loss for setups 2 to 4. While for a user in the center of the room, it may be marginally possible to exchange keys at PSD $=10^{-7}$~W/nm, once the user moves to the corner, the required PSD drops to $10^{-8}$~W/nm. This is not strange as in setups 2--4, we have more loss and additional sources of noise as compared to setup 1. The required parameter values may not, however, be achievable in practical settings, and that implies that dynamic beam steering may be needed at both the transmitter and the receiver side of a wireless QKD link.

There are several other observations that can be made from Fig.~\ref{fig_rate_vs_couping_loss_case1_2}. We have verified that the MDI-QKD with DS has a rather poor performance, and in order to tolerate substantial coupling loss, we need to use nearly ideal single-photon sources. It can also be seen that the performance of setups 3 and 4 is more or less the same. As expected, moving the BSM module around does not make a big difference in the key rate. Setup 3 has slightly better performance for the parameter values chosen here, partly because setup 4 might have slightly more Raman noise, as will be shown later. But, overall, if one needs to go with a trust-free relay node, its position can be decided based on the operational convenience without sacrificing much of the performance. In forthcoming graphs, we then only present the results for setup 3.

\begin{figure}[tb]
	\centering
	\includegraphics[width=.97\linewidth]{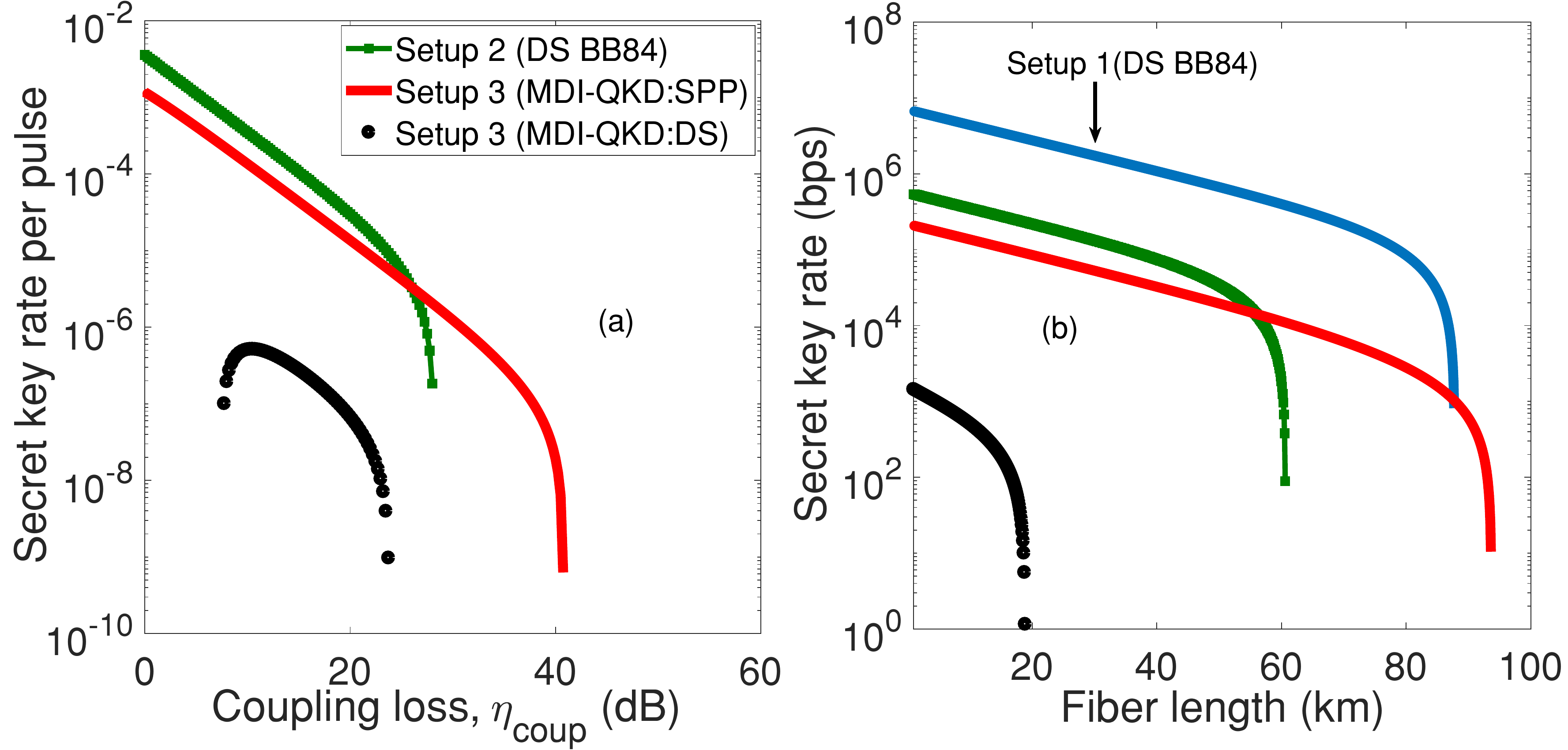}
	\caption{The secret key rate for setups 1--3 in case 3, in which the full alignment between the QKD node on the ceiling and wireless transmitter is obtained. The QKD source is placed at a corner of the room's floor, with semi-angle at half power $\Phi_{1/2}$ = 1$^{\circ}$. Receiver's FOV is 6$^{\circ}$. (a) The secret key rate per pulse versus the coupling loss, $\eta_{\rm coup}$, in dB. Fiber length is $L_0=10$ km and PSD is $10^{-5}$ W/nm. (b) The total secret key rate in bps versus $L_0$ when the coupling loss is 10 dB, PSD is $10^{-5}$ W/nm, and the repetition rate is 1~GHz. (DS: Decoy state; SPP: Single-photon pulse.)}
	\label{fig_rate_vs_couping_loss_case3}
\end{figure}

The situation is much more optimistic if full alignment, with $\Phi_{1/2}$ = 1$^{\circ}$, between the wireless QKD receiver and transmitter is attained (case 3). In this case, the QKD source is located at a corner of the room and transmits directly to the QKD receiver or the collector. The full alignment  for this narrow beam would highly improve the channel transmissivity. Figure \ref{fig_rate_vs_couping_loss_case3}(a) shows key rate versus coupling loss at a PSD of $10^{-5}$~W/nm. It can be seen that coupling loss as high as 40~dB can be tolerated in certain setups. That leaves a large budget for loss in different elements of the system. As compared to Fig.~\ref{fig_rate_vs_couping_loss_case1_2}, the rate has also improved by around three orders of magnitude. For a fixed coupling loss of 10~dB, Fig.~\ref{fig_rate_vs_couping_loss_case3}(b) shows how the remaining loss budget can be used to reach farther central offices. It seems that tens of kilometers are reachable with practical decoy-state signals in all setups. In this figure, we have also shown the total key rate for setup 1, which can serve as a benchmark for other setups. For a repetition rate of 1~GHz, keys can be exchanged at a total rate ranging from kbps to Mbps at moderate distances. 

\begin{figure}[tb]
	\centering
	\includegraphics[width=.99\linewidth]{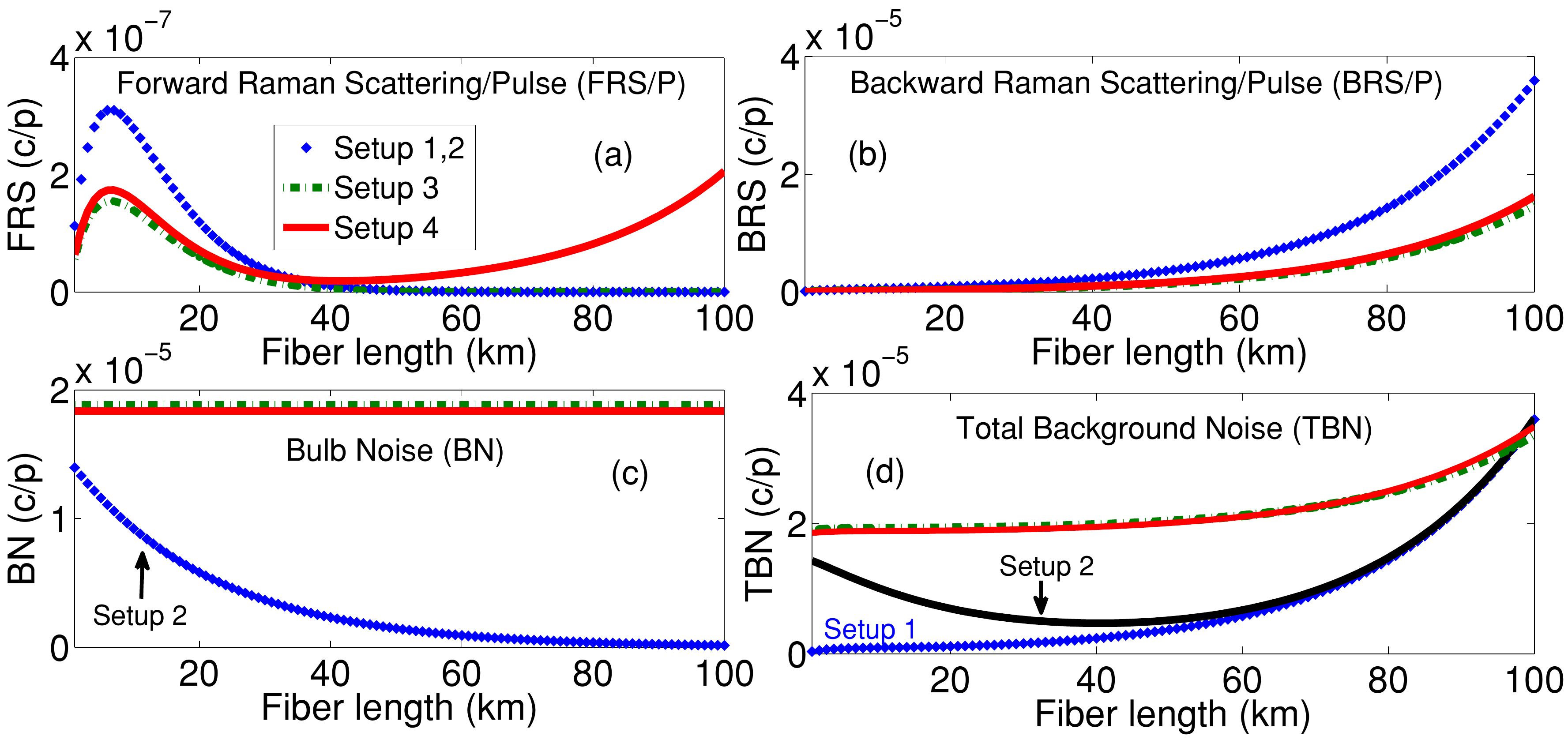}
	\caption{Noise counts per detector due to (a) forward Raman scattering, (b) backward Raman scattering, (c) the artificial lighting source, and (d) the total background noise $n_N$, all in count per pulse (c/p), versus $L_0$. The bulb's PSD is $10^{-5}$~W/nm and $\eta_{\rm coup}$ is 10 dB.}
	\label{fig_four_figures}
\end{figure}

{There are additional interesting, but somehow puzzling, points in Fig.~\ref{fig_rate_vs_couping_loss_case3}. For instance, in Fig.~\ref{fig_rate_vs_couping_loss_case3}(a), the MDI-QKD curve with DS implies that no secret keys can be exchanged at low coupling losses. This is counter-intuitive. But, we have verified that the same behavior is seen in asymmetric MDI-QKD systems, when one user's, let's say Alice, signal is accompanied by a background noise. Such a background noise would therefore undergo the same amount of loss as the Alice signal. In a particular regime, where the background noise is comparable to Bob's rate of photon arrival at the BSM module, such background photons could masquerade Bob's photons and cause errors. In setup 3, the background noise that accompanies Alice's signal is that of the bulb noise. If we make the coupling loss very low, such a noise would easily get into our BSM module and can cause errors. This explains the strange behavior of the MDI-QKD curve in Fig.~\ref{fig_rate_vs_couping_loss_case3}(a). Another detailed point is in Fig.~\ref{fig_rate_vs_couping_loss_case3}(b), in which the maximum security distance for setup 2, with 10~dB of coupling loss, is 60~km. In that case, one may expect that the security distance for setup 1, with no coupling loss should be 50~km (corresponding to 10~dB of fiber loss) longer, i.e., 110~km. The difference is, however, around 30~km. This turns out to be because of the additional Raman noise at longer distances. In order to understand this and the previous observation better, we need to explore the noise characteristic of the system, as we do next.}

{In Fig.~\ref{fig_four_figures}, we have plotted the noise counts per detector due to (a) forward Raman scattering (FRS), (b) backward Raman scattering (BRS), (c) the lighting source bulb, and (d) the total background noise $n_N$ for each setup. In each setup, the (a)--(c) noise components have been obtained from the corresponding expression for $n_N$ by breaking it into its individual terms. There are several observations to be made. In terms of order of magnitude, all three sources of noise in Figs.~\ref{fig_four_figures}(a)-(c), are larger or comparable to dark count noise per pulse, where the latter in our setup is $10^{-7}$/pulse. This proves the relevance of our analysis that accounts for Raman and background noises. In Fig.~\ref{fig_four_figures}(a), the FRS in setup 4 has a surprising rise at long distances. This is because of the launch power control scheme in use, which requires the data transmitters to send a larger amount of power proportional to the channel loss. At a short fixed $L_1$, this additional power creates additional FRS in setup 4. The effect of FRS is, however, negligible when compared to BRS, which is roughly two orders of magnitude higher than FRS. BRS increases with fiber length because of the power control scheme, and will be the major source of noise in long distances. This increase in BRS justifies the shorter-than-expected security distances in Fig.~\ref{fig_rate_vs_couping_loss_case3}(b). Finally, it can be seen that why MDI-QKD setups are more vulnerable to bulb noise than the DS system of setup 2. The bulb noise would enter the BSM module in setups 3 and 4 by mainly being attenuated by the coupling loss, whereas in setup 2, it will be further attenuated by the channel loss. That is partly why the rate in setup 2 can be higher than that of setups 3 and 4. Based on these results, one can conclude that, if the MDI property is not a crucial design factor, setup 2 could offer a reasonable practical solution to the scenarios where a trusted relay is not available. In the rest of this section, we will then compare the performance of different protocols that can be run in setup 2.}

\begin{figure}[tb]
	\centering
	\includegraphics[width=.97\linewidth]{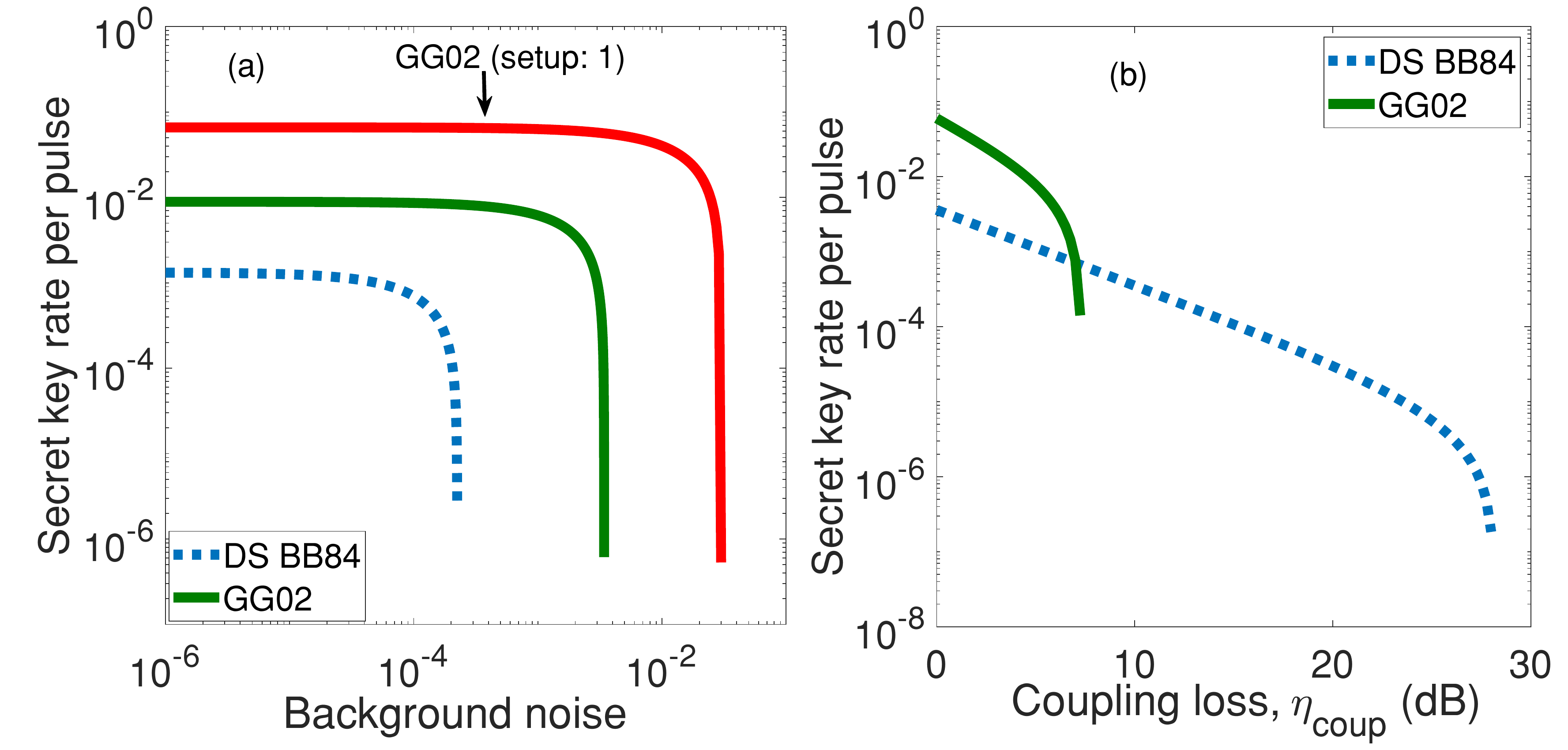}
	\caption{Comparison of the GG02 and DS-BB84 protocols for setup 2 and case 3 (except for the curve labeled GG02 (setup 1)). (a) Secret key rate per pulse versus total background noise. The latter is assumed to be per detector for DV-QKD, while it is per spatio-temporal mode for CV-QKD. (b) Secret key rate per pulse versus coupling loss, $\eta_{\rm coup}$, in dB. The coupling loss in (a) is 5 dB for setup 2 and 0 dB for setup 1. The shared fiber length ($L_0$) is 10 km. The used bulb's PSD is $10^{-5}$ W/nm.}
	\label{fig_three_figures_case3}
\end{figure}

Figure \ref{fig_three_figures_case3} compares the GG02 performance in setups 1 and 2 with DS-BB84. In Fig.~\ref{fig_three_figures_case3}(a) we study the resilience of either scheme against background noise at low values of coupling loss. As has been shown for fiber-based systems \cite{CV_ResilienceBQi}, CV-QKD can tolerate a higher amount of background noise in this regime due to the intrinsic filtering properties of its local oscillator. That benefit would however go away if the coupling loss roughly exceeds 10~dB in our case; see Fig.~\ref{fig_three_figures_case3}(b). {This implies that full beam steering is definitely a must when it comes to CV-QKD.} Depending on the setting of the system, the operator can decide whether a DV or a CV scheme is the better option.

\begin{figure}[tb]
	\centering
	\includegraphics[width=.97\linewidth]{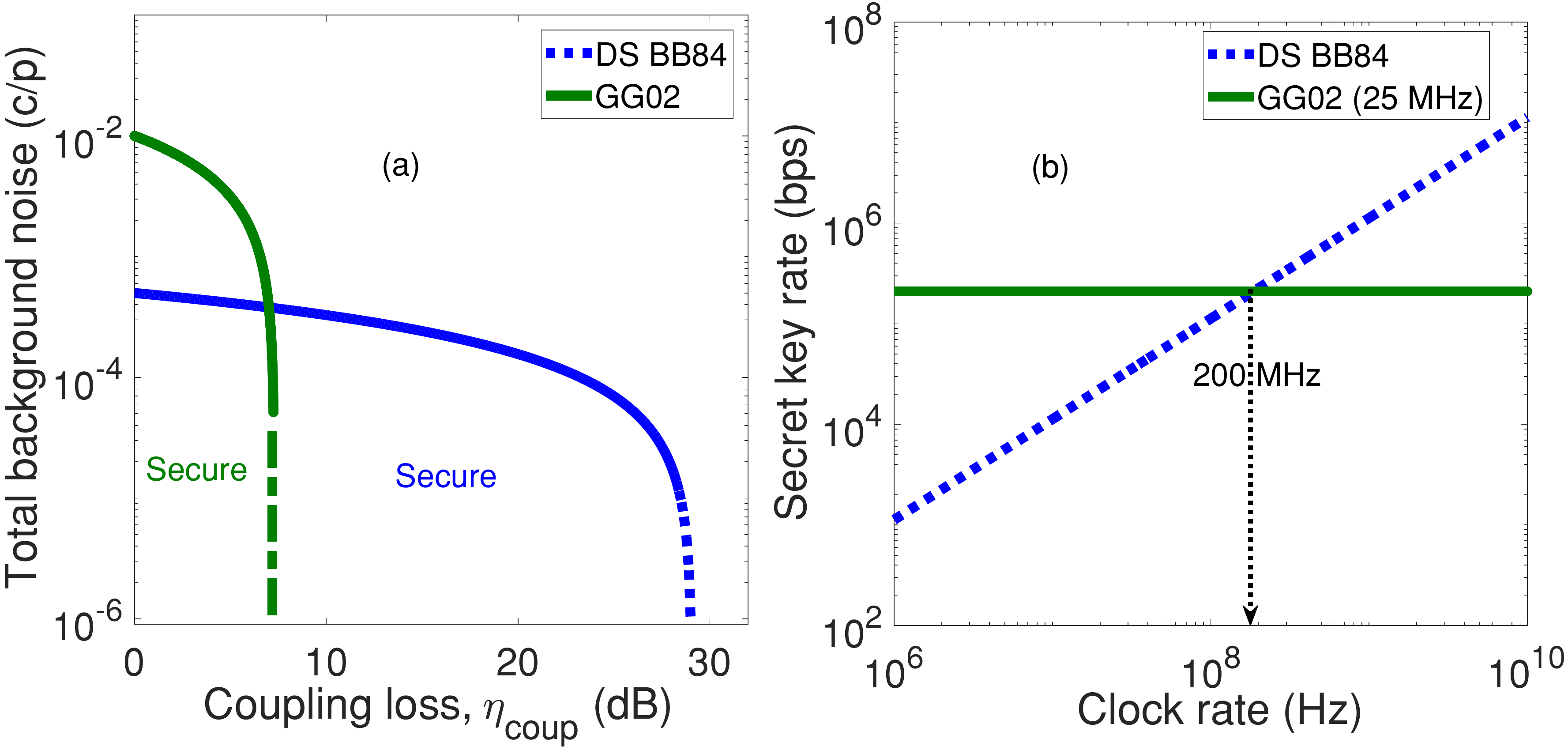}
	\caption{ (a) Regions of secure operation for DV-QKD (DS-BB84) and CV-QKD (GG02) protocols for setup 2 (case 3). The curves show the maximum tolerable background noise at different values of coupling loss, $\eta_{\rm coup}$, in dB. The background noise is calculated per detector for DV-QKD, while it is per spatio-temporal mode for CV-QKD. (b)  Comparison of  the two systems from the clock rate point of view when the CV repetition rate is fixed to 25 MHz. In (a) and (b), $L_0 = 10$~km. In (b), coupling loss is 5 dB and PSD is $10^{-5}$~W/nm.}
	
	\label{fig_tradeoff}
\end{figure}

Figure~\ref{fig_tradeoff} shows the relevant regimes of operation for DV and CV-QKD schemes in a different way. In Fig.~\ref{fig_tradeoff}(a), we have looked at the maximum coupling loss tolerated by each of the two schemes for a given background noise. It is clear that while for low values of coupling loss, CV-QKD can tolerate more noise, at high values of coupling loss DV-QKD is the only option, although it can tolerate less noise. There is therefore a trade-off between the amount of coupling loss versus background noise the system can tolerate. In Fig.~\ref{fig_tradeoff}(b), we have compared the two systems from the clock rate point of view. CV-QKD is often practically constrained by its low repetition rate. In Fig.~\ref{fig_tradeoff}(b), we have fixed the CV repetition rate to 25 MHz~\cite{wang201525} and have found out at what clock rate the DV system offers a higher total key rate than the CV one. For numerical values used in our simulation this cross-over rate is around {200}~MHz, which is achievable for today's DV-QKD systems. The ultimate choice between DV and CV would then depend on the characteristics of the system, such as loss and noise levels, as well as the clock rate available to the QKD system.


\section{Conclusions}
\label{Sec:Conclusions}
We proposed and studied four configurations that enabled wireless access to hybrid quantum-classical networks. All these setups included an initial wireless indoor link that connected a quantum user to the network. Each user, in the access network, could also communicate classically with the central office via another wavelength in the same band. We considered setups in which a local relay point could be trusted, as well as setups where such a trust was not required. We showed that with proper beam alignment it was possible, in both DV- and CV-QKD, to achieve positive key rates for both trusted and untrusted relay points in certain indoor environments. 

{The choice of the optimum setup would depend on various system parameters, which we studied in our analysis. For instance, we found that our MDI-QKD setups, which offered trust-free QKD immune to measurement attacks, were mostly insensitive to the position of their measurement module, but could suffer harshly from the background noise generated in the indoor environment. If the immunity to measurement attacks was not required, we could simply collect QKD signals at the ceiling and couple them into optical fibers along with other data channels. With decoy-state techniques, we showed that we could tolerate up to 30~dB of coupling loss in such a setting, provided that full alignment is achieved. At long distances, the Raman noise induced by the data channels would also take its toll on the maximum secure distance limiting it to tens of kilometers. Both Raman noise and the background noise due to the artificial light source in the indoor environment could be orders of magnitude larger than the static dark count of single-photon detectors. We also showed that in the low coupling loss regime, CV-QKD could offer higher rates and more resilience to background noise than DV-QKD systems. But, overall, DV-QKD schemes could offer a more stable and flexible operation adaptable to a wider range of scenarios. In short, using our analytical results, we can identify the winner in realistic setups that enable high-rate wireless access to future quantum networks.} 

\appendices
\section{DS-BB84 key rate analysis}
\label{App:DS-BB84_KeyRate}

In this appendix, the secret key generation rate of the DS-BB84 protocol is calculated. The lower bound for the key rate, in the limit of an infinitely long key, is given by~\cite{MXF:Practical:2005}
\begin{align}
\label{App:KeyRateLB}
R\geq q\lbrace -Q_\mu f h(E_\mu)+Q_1[1-h(e_1)]\rbrace, 
\end{align}
where $q$ is the basis-sift factor, which is assumed to approach 1 in the efficient BB84 protocol \cite{Lo2005efficient} as employed in this work. The error correction inefficiency is denoted by $f>1$ and $\mu$ is the average number of photons per signal pulse. Moreover, in \eqref{App:KeyRateLB}, $Q_\mu$, $E_\mu$, $Q_1$, $e_1$ and $h(x)$ are, respectively, the overall gain, the quantum bit error rate (QBER), the single-photon gain, the error rate in single-photon states and the Shannon binary entropy function. In the case of a lossy channel with a total transmissivity of $\eta$ and a total background noise per detector of $n_N$, the above parameters are given by~\cite{Panayi2014memory}:
\begin{align}
Q_\mu=& 1-e^{-\eta\mu}(1-n_N){^2}, \nonumber \\
E_{\mu}=& \frac{e_{0}{Q_\mu}-({e_0}-{e_d})(1-e^{-\eta\mu})(1-n_N)}{Q_\mu}, \nonumber \\
Q_1=& Y_{1}\mu{e^{-\mu}}, e_1= \frac{e_0{Y_1}-({e_0}-{e_d})\eta(1-n_N)}{Y_1}, 
\end{align}
where $e_0=1/2$ and  
\begin{align}
Y_1=& 1-(1-\eta)(1-n_N)^2, \nonumber \\
h(x)=& -x\log{_2}x-(1-x)\log{_2}(1-x),
\end{align}
where we assume that there has been no eavesdropping activity in the channel. This is considered to be the normal oeprating mode of the system and the key rate calculated under above conditions would give us a sense of what we may expect from our QKD system in practice. The same assumptions have been used to calculate the key rate of other protscols as we see next.

\section{GG02 key rate analysis}
\label{App:GG02_KeyRate}
The secret key rate for GG02 with reverse reconciliation, under collective attacks, is given by~\cite{fossier2009field}
\begin{align}
\label{App:GG02_KeyRateFormula}
K= \beta I_{AB}-\chi_{BE}, 
\end{align}
where $\beta$ is the reconciliation efficiency, $I_{AB}$ is the mutual information between Alice and Bob, which, for a Gaussian channel, is given by
\begin{align} 
I_{AB}= \frac{1}{2} \log_{2}\frac{V+\chi_{tot}}{1+\chi_{tot}},
\end{align}
where $V$ and $\chi_{tot}$ are, respectively, the total variance and the total noise given by
\begin{align} 
V= V_{A}+1,
\end{align}
with $V_A$ being the variance of Alice's quadrature modulation and
\begin{align} 
\chi_{tot}=\chi_{line}+\chi_{hom}/\eta_{\rm ch},
\end{align}
in which 
\begin{align} 
\label{channel_noise}
\chi_{line}=& \frac{1-\eta_{\rm ch}}{\eta_{\rm ch}} +\varepsilon, \nonumber \\
\chi_{hom}=& \frac{1-\eta_{B}}{\eta_{B}} +\frac{v_{elec}}{\eta_{B}},
\end{align}
are, respectively, the noise due to the channel and the noise stemming from homodyne detection. Also, the parameters $\eta_{B}$, $v_{elec}$, $\varepsilon$ and $\eta_{\rm ch}$, are, respectively, Bob's overall efficiency, electronic noise variance induced by homodyne electronic board, excess noise, and the channel transmittance. 

In \eqref{App:GG02_KeyRateFormula}, $\chi_{BE}$ is the Holevo information between Eve and Bob, and it is given by
\begin{align}
\chi_{BE}= g(\Lambda_1)+g(\Lambda_2)-g(\Lambda_3)-g(\Lambda_4), 
\end{align}
where
\begin{align}
g(x)=(\frac{x+1}{2})\log_{2}(\frac{x+1}{2})-(\frac{x-1}{2})\log_{2}(\frac{x-1}{2}),
\end{align}
with 
\begin{eqnarray} 
&\Lambda_{1/2}= {\sqrt{(A\pm \sqrt{A^2-4B})/2}}, &\nonumber \\
&\Lambda_{3/4}= {\sqrt{(C\pm \sqrt{C^2-4D})/2}}.&
\end{eqnarray}
In the above equations:
\begin{align} 
A=& V^2(1-2\eta_{\rm ch})+2\eta_{\rm ch}+\eta_{\rm ch}^2(V+\chi_{line})^2,  \nonumber \\
B=& \eta_{\rm ch}^2(V\chi_{line}+1)^2,  \nonumber \\
C=& \frac{V\sqrt{B}+\eta_{\rm ch}(V+\chi_{line})+A\chi_{hom}}{\eta_{\rm ch}(V+\chi_{tot})}, \nonumber \\
D=& \sqrt{B}\frac{V+\sqrt{B}\chi_{hom}}{\eta_{\rm ch}(V+\chi_{tot})}.
\end{align} 

\section{MDI-QKD key rate analysis}
\label{App:MDI-QKD}
In this appendix, we summarize the secret key rate of the MDI-QKD protocol. The rates for the ideal single-photon source and the decoy-state protocols, respectively, are
\begin{align}
\label{SPP-MDIQKDrate:App}
R_{\rm MDI-QKD}^{\rm{SPP}}= Y_{11}[1-h(e_{11:X})-fh(e_{11:Z})]
\end{align}
and
\begin{align}
\label{DS-MDIQKDrate:App}
R_{\rm MDI-QKD}^{\rm{DS}}= Q_{11}(1-h(e_{11;X}))-fQ_{\mu\nu;Z}h(E_{\mu\nu;Z}).
\end{align}
In the above, $Q_{11}$ is the gain of the single-photon states given by
\begin{align}
Q_{11}=\mu\nu e^{-\mu-\nu} Y_{11},
\end{align}
where $\mu$ ($\nu$) is the mean number of photons in the signal state sent by Alice (Bob) and $Y_{11}$ is the yield of the single-photon states given by
\begin{align}
Y_{11}= & (1-n_N)^2[\eta_a \eta_b/2+(2\eta_a+2\eta_b-3\eta_a \eta_b)n_N  \nonumber \\
& +4(1-\eta_a)(1-\eta_b)n_N^2] ,
\end{align}
where $n_N$ represents the total noise per detector and $\eta_a$ and $\eta_b$ are, respectively, the total transmittance between Alice and Bob sides and that of Charlie~\cite{Panayi2014memory}. In \eqref{SPP-MDIQKDrate:App} and \eqref{DS-MDIQKDrate:App}, $e_{11;Z}$, $e_{11;X}$, $Q_{\mu\nu;Z}$ and $E_{\mu\nu;Z}$, respectively, represent the QBER in the $Z$ basis for single-photon states, the phase error for single-photon states, the overall gain and the QBER in the $Z$-basis, which are given by~\cite{Panayi2014memory}:
\begin{align}
e_{11;X}Y_{11}=& Y_{11}/2-(1/2-e_d)(1-n_N)^2 \eta_a \eta_b/2, \nonumber\\
e_{11;Z}Y_{11}=& Y_{11}/2-(1/2-e_d)(1-n_N)^2 (1-2n_N) \eta_a \eta_b/2, \nonumber\\
Q_{\mu \nu;Z}=& Q_C+Q_E,	E_{\mu \nu;Z}Q_{\mu \nu;Z}= e_dQ_c+(1-e_d)Q_E,
\end{align}
where
\begin{align}
Q_C= & 2(1-n_N)^2e^{-\mu^{'}/2}[1-(1-n_N)e^{-\eta_a \mu/2}] \nonumber \\
& \times [1-(1-n_N)e^{-\eta_b \nu/2}] \nonumber \\
Q_E=& 2n_N(1-n_N)^2e^{-\mu^{'}/2}[I_0(2x)-(1-n_N)e^{-\mu^{'}/2}],
\end{align}
with $x= \sqrt{\eta_a \mu \eta_b \nu}/2$, $\mu^{'}=\eta_a \mu+\eta_b \nu$ and $I_0$ being the modified Bessel function. 

\bibliographystyle{IEEEtran}

\bibliography{Master1}

\begin{IEEEbiographynophoto}{Osama Elmabrok}
received his B.Sc degree in electrical and electronic engineering from University of Benghazi (formerly known as Garyounis University) in 2002, and  M.Eng in communication and computer engineering from National University of Malaysia in 2007. He was working as an assistant lecturer for University of Benghazi before joining University of Leeds where is currently working toward his Ph.D. degree. Prior joining the academic field, he worked as a communications engineer for Almadar Telecom Company, Azzaawiya Refining Company, and RascomStar-QAF Company. His current research interest is about wireless quantum key distribution (QKD) in indoor environments.
\end{IEEEbiographynophoto}

\begin{IEEEbiographynophoto}{Masoud Ghalaii}
	was born in April 1987. He received the B.Sc. degree in Chemical Engineering from Isfahan University of Technology, Isfahan, Iran, in September 2011. He then pursued his studies and received the M.Sc. degree with honors in Physics with the group Quantum Information Science at Sharif University of Technology, Tehran, Iran, in January 2014. Since October 2015, he has been working toward the Ph.D. degree at the School of Electronic and Electrical Engineering at the University of Leeds, Leeds, United Kingdom. His research interests include quantum optical communications, and application of quantum amplifiers and repeaters in continuous-variable quantum key distribution, as well as quantum information science.
\end{IEEEbiographynophoto}

\begin{IEEEbiographynophoto}{Mohsen Razavi}
 received his B.Sc. and M.Sc. degrees (with honors) in Electrical Engineering from Sharif University of Technology, Tehran, Iran, in 1998 and 2000, respectively. From August 1999 to June 2001, he was a member of research staff at Iran Telecommunications Research Center, Tehran, Iran, working on all-optical CDMA networks and the possible employment of optical amplifiers in such systems. He joined the Research Laboratory of Electronics, at the Massachusetts Institute of Technology (MIT), in 2001 to pursue his Ph.D. degree in Electrical Engineering and Computer Science, which he completed in 2006. He continued his work at MIT as a Post-doctoral Associate during Fall 2006, before joining the Institute for Quantum Computing at the University of Waterloo as a Post-doctoral Fellow in January 2007. Since September 2009, he is a Faculty Member at the School of Electronic and Electrical Engineering at the University of Leeds. His research interests include a variety of problems in classical optical communications. In 2014, he chaired and organized the first international workshop on Quantum Communication Networks. He is the Coordinator of the European project QCALL, which endeavors to make quantum communications technologies available to end users.
\end{IEEEbiographynophoto}

\end{document}